\documentclass[useAMS,usenatbib]{mn2e}
\usepackage{epsfig}
\usepackage{times}
\usepackage{color}

\def\cm3{cm$^{-3}$}
\def\kms{km~s$^{-1}$}

\def\lsun{L$_{\odot}$}

\def\msun{M$_{\odot}$}

\def\beq{\begin{equation}}
\def\eeq{\end{equation}}

\def\lesssim{\mathrel{\hbox{\rlap{\hbox{\lower4pt\hbox{$\sim$}}}\hbox{$<$}}}}
\def\gtrsim{\mathrel{\hbox{\rlap{\hbox{\lower4pt\hbox{$\sim$}}}\hbox{$>$}}}}

\def\aj{AJ}
\def\pasp{PASP}
\def\apj{ApJ}
\def\apjs{ApJS}
\def\apjl{ApJL}
\def\aap{A\&A}

\def\mnras{MNRAS}
\def\nat{Nature}
\def\iaucirc{IAU~Circ.}

\definecolor{red}{rgb}{1,0,0}

\title[Radiation hydrodynamics of explosions/eruptions in massive stars]{Shock-heating of stellar envelopes:
A possible common mechanism at the origin of explosions and eruptions in massive stars}

\author[Luc Dessart, Eli Livne, and Roni Waldman]
{
Luc Dessart$^{1}$\thanks{E-mail: Luc.Dessart@oamp.fr},
Eli Livne$^{2}$, and Roni Waldman$^{2}$\\
$^{1}$ Laboratoire d’Astrophysique de Marseille, Universit\'e de Provence,
CNRS, 38 rue Fr\'ed\'eric Joliot-Curie, F-13388 Marseille Cedex 13, France \\
$^{2}$ Racah Institute of Physics, The Hebrew University, Jerusalem, Israel
}

\begin{document}

\date{Accepted 2010 February 28. Received 2010 February 19; in original form 2009 October 19}

\pagerange{\pageref{firstpage}--\pageref{lastpage}} \pubyear{2009}

\maketitle

\label{firstpage}

\begin{abstract}
 Observations of transient phenomena in the Universe reveal a spectrum of mass-ejection properties
 associated with massive stars, covering from Type II/Ib/Ic core-collapse supernovae (SNe)
 to giant eruptions of Luminous Blue Variables (LBV) and optical transients. In this work,
 we hypothesize that a large fraction of these phenomena may have an explosive origin, the distinguishing ingredient
 being the ratio of the prompt energy release $E_{\rm dep}$ to the envelope binding energy $E_{\rm binding}$.
 Using one-dimensional one-group radiation hydrodynamics and a set of 10-25\,\msun\, massive-star models,
 we explore the dynamical response of a stellar envelope subject to a strong, sudden, and deeply-rooted energy release.
 Following energy deposition, a shock {\it systematically} forms, crosses the
progenitor envelope on a day time-scale,
 and breaks-out with a signal of hour-to-days duration and a 10$^5$-10$^{11}$\,\lsun\, luminosity.
 We identify three different regimes, corresponding to a transition from dynamic to quasi-static diffusion transport.
 For $E_{\rm dep} > E_{\rm binding}$, full envelope ejection results with a SN-like bolometric luminosity and kinetic energy,
 modulations being commensurate to the energy deposited and echoing the diversity of Type II-Plateau SNe.
 For $E_{\rm dep} \sim E_{\rm binding}$, partial envelope ejection results with a small expansion speed, and a more modest but
 year-long luminosity plateau, reminiscent of LBV eruptions or so-called SN impostors.
 For $E_{\rm dep} < E_{\rm binding}$, we obtain a ``puffed-up'' star, secularly relaxing back to thermal equilibrium.
 In parallel with gravitational collapse and Type II SNe, we argue that thermonuclear combustion, for example of as little as
 a few 0.01\,\msun\, of C/O, could power a wide range of explosions/eruptions.
 Besides massive stars close to the Eddington limit and/or critical rotation, 8-12\,\msun\, red-supergiants,
 which are amongst the least bound of all stars, represent attractive candidates for transient phenomena.
\end{abstract}

\begin{keywords} radiation hydrodynamics -- stars: atmospheres -- stars:
supernovae - stars: transients
\end{keywords}

\section{Introduction}
\label{sect_intro}

      Stellar explosions, broadly refered to as supernovae (SNe), are understood to stem from a sudden
release of energy either associated with the collapse of the degenerate core of a massive star or
from the thermonuclear combustion of fresh fuel deep inside the stellar envelope.
Whether one or the other mechanism occurs seems to depend on the main-sequence mass of the progenitor star,
with core collapse occuring systematically if its value is above $\sim$8\,\msun\, \citep[hereafter WHW02]{WHW02}.
Interestingly, whatever the mechanism, the typical kinetic energy of SN ejecta is on the order of 10$^{51}$\,erg,
as inferred for example for the well-studied SN 1987A (Type II peculiar; \citealt{blinnikov_etal_2000}),
for SN 1999em (Type II-Plateau, heareafter II-P; \citealt{utrobin_07}), or for the very uniform set of events that
Type Ia SNe constitutes \citep{WKB07_Ia_lc}. The cause of this apparent degeneracy in explosion energy is,
paradoxically, perhaps not so much tied to the mechanism itself, but instead
to the rather uniform total envelope binding energy of the progenitor stars, on the order of 10$^{51}$\,erg;
anything falling short of that leads to a fizzle and no SN display.

\begin{figure*}
\epsfig{file=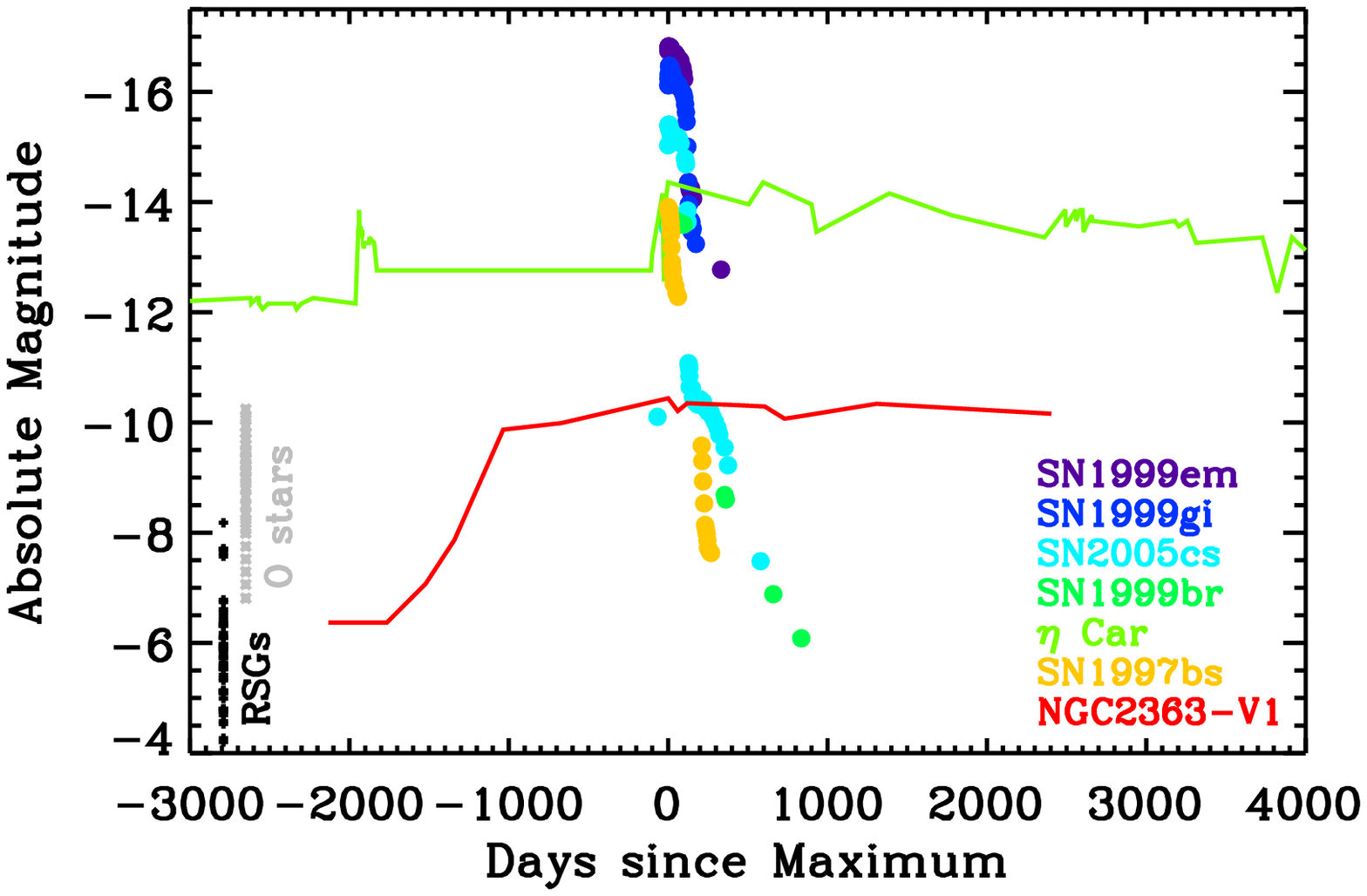,width=8.5cm}
\epsfig{file=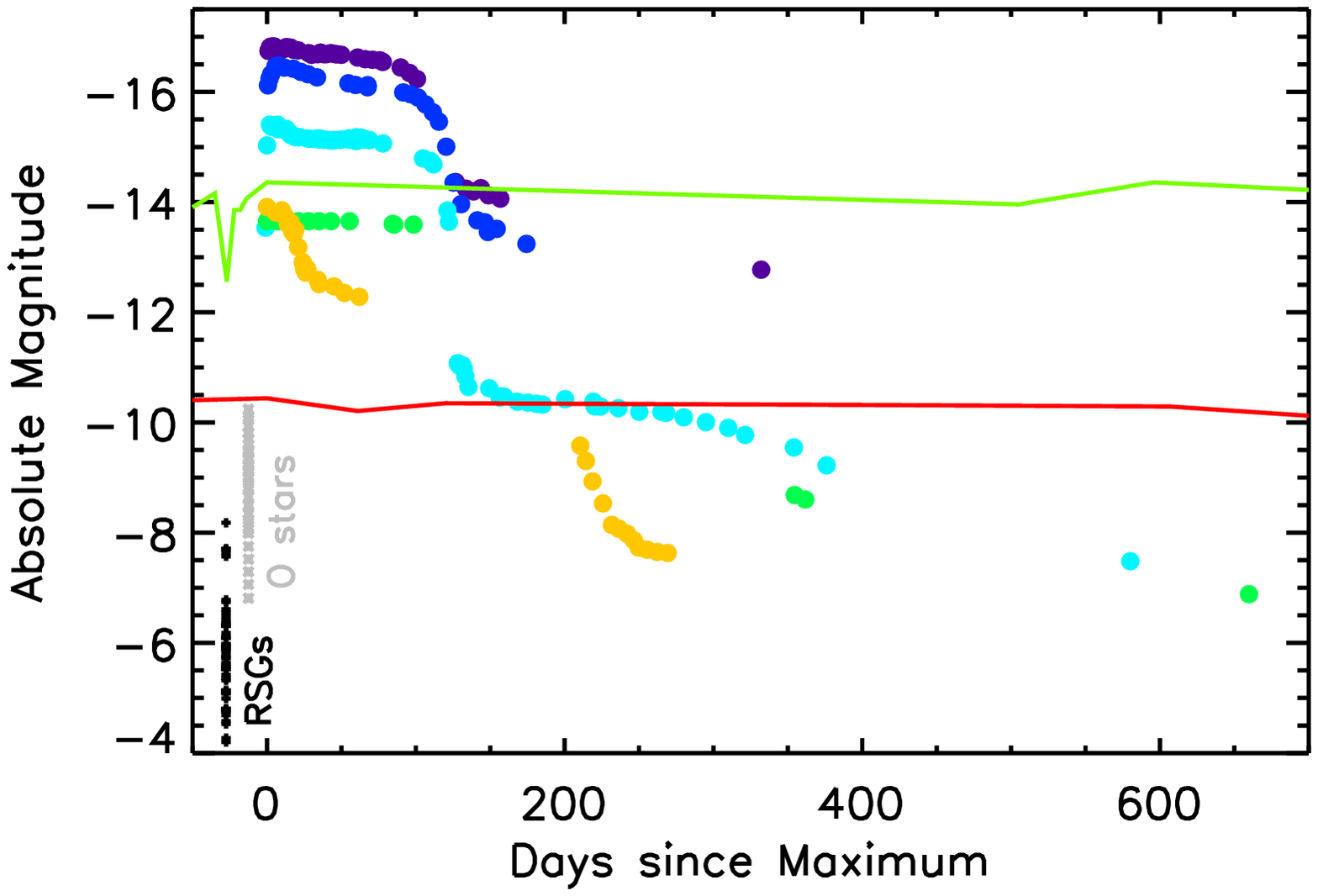,width=8.5cm}
\caption{{\it Left:} Comparison of absolute-$V$-band-magnitude light curves for an
illustrative and non-exhaustive sample of SNe, SN impostors and/or erupting LBVs
(violet: SN1999em, \citealt{leonard_etal_02a,DH06_SN1999em}; blue: SN1999gi, \citealt{leonard_etal_02b};
turquoise: SN 2005cs, \citealt{pastorello_etal_09}; green: SN1999br, \citealt{pastorello_etal_09}:
light green: $\eta$ Car, \citealt{frew_04}; yellow: SN 1997bs, \citealt{vandyk_etal_00};
red: NGC2363-V1, \citealt{drissen_etal_01,petit_etal_06}).
For each, we adopt the distance and reddening given
in the associated references. The time origin is that of maximum recorded brightness.
Note that all these objects have comparable effective temperatures on the order of 10,000\,K, hence
comparable bolometric correction, making the comparison of their absolute $V$-band magnitude meaningful.
We also show on the left side and in black the absolute visual magnitude of the galactic red-supergiant
stars studied by \citet[black]{levesque_etal_05}, as well as the {\it bolometric} magnitude
of the O star models computed by \citet[gray; we use the bolometric magnitude here since
O stars are hot and have large bolometric corrections]{martins_etal_05}.
{\it Right:} Same as left, but now zooming in on the time of maximum brightness (the color coding is the same
as in the left panel).
Notice the stark contrast between SN light curves, associated with shorter/brighter events, and erupting massive stars,
associated with longer/fainter events. Importantly, notice the overlap between the intrinsic brightness of $\eta$ Car and that of
the low-luminosity Type II-P SN 1999br. In this work, we propose that this diversity of radiative displays may be
accomodated by a {\it common, explosive, origin}.
\label{fig_obs_lc}
}
\end{figure*}

     The last decade of observations of such transient phenomena has shown, however, that the radiative signatures
associated with SNe (as classified in circulars) are very diverse, from
very faint to very luminous, from fast-evolving to slow-evolving or fast-expanding to slow-expanding.
This diversity has been observed little in Type Ia SNe, with a few peculiar events such as
SN 2002ic (presence of narrow hydrogen lines in an otherwise standard Type Ia spectrum; \citealt{hamuy_iau_02ic})
or SNLS-03D3bb (possible Type Ia SN from a super-Chandrasekhar white dwarf star; \citealt{howell_etal_06}).
In contrast, there has been a rich diversity in explosions associated (perhaps erroneously at times) with massive
stars and the mechanism of core collapse. We have observed 1) Type Ic SNe, associated or not with a long-soft $\gamma$-ray burst,
and with a standard or a very large kinetic energy (SN 1998bw, \citealt{WES_99}; SN 2002ap, \citealt{mazzali_etal_02});
2) a large population of Type II-Plateau (II-P) SNe in what seems to be the
generic explosion of a moderate-mass red-supergiant (RSG) star (e.g. SN 2005cs, \citealt{maund_etal_05,UC_08});
3) a growing number of low-luminosity SNe that share properties with standard Type II-P SNe except for being
significantly and globally less energetic (e.g. SN1997D, \citealt{chugai_utrobin_00};
SN 1999br, \citealt{pastorello_etal_04}; OT2006-1 in M85, whose status
is ambiguous, see \citealt{kulkarni_etal_2007_m85,pastorello_etal_07_m85}).
We show a sample of $V$-band absolute-magnitude light curves of such core-collapse SNe in Fig.~\ref{fig_obs_lc},
with representative peak values  of $-$14 to $-$17\,mag and a 100-day plateau duration (best seen in the right panel
of that figure), hence about 6-10\,mag brighter than their proposed RSG progenitors (shown as black crosses).
From this expanded SN sample, the range of corresponding explosion energies has considerably widened, extending above and below
the standard 10$^{51}$\,erg value. Within the core-collapse SN context, this modulation is thought to stem
from modulations in the energy revival of the stalled shock above the nascent
proto-neutron star, in turn modulated by the stellar-core structure \citep{burrows_etal_07a}.

    The stretching to low explosion energies of potential core-collapse SN events is intriguing.
For the most energetic explosions belonging to points 1 and 2 above, the classification as a SN is unambiguous.
However, some transient events show an ejecta/outflow kinetic energy and a peak magnitude that are SN-like, although
the events did not stem from core collapse (a star is observed at that location on post-explosion/eruption images);
the community calls these SN impostors (e.g. SN1997bs; Fig.~\ref{fig_obs_lc}; \citealt{vandyk_etal_00}). 
Conversely, this raises the issue whether {\it low-energy} Type II-P SNe are associated with core collapse - they might but
they need not.
We illustrate this overlap in radiative properties for a sample of such objects in Fig.~\ref{fig_obs_lc}.
One such case is the Luminous Blue Variable (LBV) $\eta$ Car, whose properties
during its 1843 eruption rival those of the low-luminosity Type II-P SN1999br.
$\eta$ Car survived this gigantic eruption, which shed about 10\,\msun\, of material
in what now constitutes the homunculus nebula \citep{smith_etal_2003}.
In contrast to core-collapse SNe, such eruptive phenomena in massive stars have been
associated with the proximity of the star to the Eddington luminosity $L_{\rm Edd} = 4 \pi c G M / \kappa$
($\kappa$ is the mass absorption coefficient). Due to the steep dependence of luminosity
to mass (e.g. with an exponent of 3.5 for main-sequence objects), this limit is easily reached by very massive
stars such as $\eta$ Car, or more generally massive blue-supergiant stars.
In this context, massive stars are thought to undergo considerable mass loss when their luminosity
overcomes the Eddington limit,\footnote{Note, however, that energy will have to be supplied to the stellar
envelope to push it over this limit, and in large amounts to explain such a nebula as the homunculus.}
giving rise to a porosity-modulated continuum-driven outflow \citep{shaviv_00,owocki_etal_04}.
Here, this super-Eddington wind constitutes a quasi steady-state outflow, and has therefore been thought to be
of a fundamentally different nature from core-collapse SN ejecta. And indeed, one refers to a wind for the former
and to an ejecta for the later. This dichotomy has been exacerbated by the stark contrast in typical light curves
of eruptive stars (long lived with large brightness) and core-collapse SN explosions (short lived with huge brightness).
In Fig.~\ref{fig_obs_lc}, we show two known eruptive massive stars that highlight this contrast.

   However, recent observations may be challenging such a strict segregation.
   First, the recent identification of very fast outflowing material ahead of $\eta$ Car's homunculus now suggests
   that such material was accelerated by a shock, rather than driven in a quasi-steady wind,
   and thus connects the giant outburst to an explosive origin \citep{smith_08_blast}.
   Second,  the existence of interacting SNe tells us that a massive eruption can occur merely a
   few years before explosion. For some, e.g.
   SN 2006gy \citep{smith_etal_07a,smith_etal_07b, woosley_etal_07} or SN 1994W \citep{dessart_etal_09},
   the amount is thought to be large enough to decelerate the energetic (and necessarily faster-expanding) subsequent
   ejection.
   This very strict timing of merely a few years, {\it which is orders of magnitude smaller than evolutionary or transport
   time-scales}, suggests a connection between the mechanisms at the origin of the two ejections.
   For SN2006gy, Woosley et al. propose recurrent pair-instability pulsations, a mechanism germane to super-massive
   stars and therefore extremely rare. For lower mass massive stars,
   this short delay of a few years seems to exclude a very-long, secular, evolution for the
   production of the first ejection since this would have no natural timing to the comparatively instantaneous event of core collapse.
   Motivated by these recent observations, we explore in this paper whether this diversity of events could be reproduced by
   a unique and deeply-rooted mechanism, associated with the sudden energy release above the stellar core and
   the subsequent shock heating of the progenitor envelope. This means would communicate a large energy to the
   stellar envelope on a shock crossing time-scale of days rather than on a very long-diffusion time-scale of thousands of years or more.
   Although different in their origin, this energy release could be a weak analogue of what results in pair-instability pulsations,
   i.e. a nuclear flash, as identified in the 8-12\,\msun\, range by \citet{weaver_woosley_79}.

   In this paper, following this shock-heating hypothesis, we use 1D radiation-hydrodynamics simulations
   to explore the production of explosions/eruptions
in stars more massive that $\sim$10\,\msun\, on the main sequence. Rather than focusing on specific models, like
those potentially associated with failed supernovae \citep{fryer_etal_09}, we parameterize the problem through a simple energy
deposition, taking place with a given magnitude, over a given time, and at a given depth in a set of pre-SN progenitor star models.
We do not aim at reproducing any specific observation but, through a systematic approach, try to identify important trends,
in a spirit similar to that of \citet{falk_arnett_77}. However, we depart from these authors by studying
``non-standard''  explosions. In practice, we
consider cases where the energy deposited can be both smaller or larger than the binding energy of the overlying envelope,
but must imperatively be released on a very short time-scale to trigger the formation of a shock.
Doing so, we identify three regimes, with ``standard'' SN explosions (short-lived transients) at the high energy end,
objects that we will group in the category SN ``impostors'' (long-lived transients) at intermediate energy,
and variable stars at the very low energy end.
 Let us stress here that we do not make the claim that all massive-star eruptions, or all transients in general, stem from a strong, sudden,
and deeply-rooted energy release in their envelope. Here, we make this our working hypothesis and investigate
how much such an explosive scenario can explain observations. We do find that this scenario has great potential
and should be considered as a possibility when examining the origin of massive-star
eruptions and associated transient phenomena.

\begin{figure}
\epsfig{file=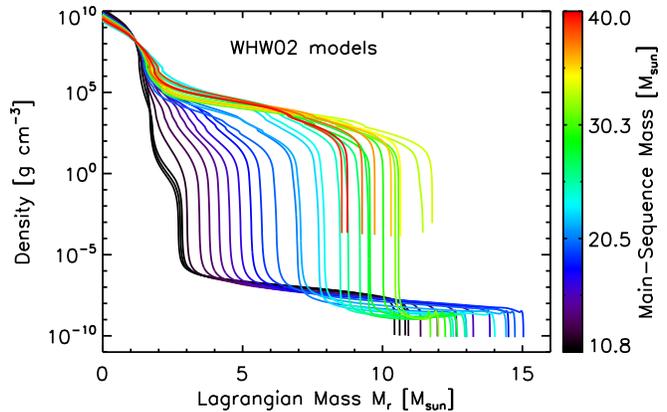,width=8.5cm}
\caption{Density distribution as a function of Lagrangian mass at the onset of collapse
for the models of WHW02 evolved at solar metallicity. Notice the flattening density distribution
above the core for increasing mass. In low-mass massive stars, the star is structured as a dense
inner region (the core), and a tenuous extended H-rich envelope.
\label{fig_rho_mr}
}
\end{figure}

   The paper is structured as follows.
   In \S\ref{sect_input}, we briefly present the stellar evolutionary models of WHW02 that we use
as input for our 1D 1-group radiation-hydrodynamics simulations. We discuss the properties of
the progenitor massive stars of WHW02, such as density structure and binding energy, that are relevant for the present study.
We then describe in \S\ref{sect_model}, the numerical technique and setup for our energy deposition study.
In \S\ref{sect_s11}, we present the results of a sequence of simulations based primarily on the 11\,\msun\, model
of WHW02, discussing the properties of the shocked progenitor envelope for different values of the strength (\S\ref{var_edep}),
the depth (\S\ref{var_mcut}) and the duration (\S\ref{var_dt}) of the energy deposition. We also discuss in \S\ref{var_mprog}
the results obtained for more massive pre-SN progenitors, ranging from 15 to 25\,\msun\, on the main sequence.
In \S\ref{sect_rad},
we present synthetic spectra computed separately with CMFGEN \citep{HM98_lb,DH05_qs_SN,dessart_etal_09}
for a few models at a representative time after shock breakout.
In \S\ref{sect_discussion}, we discuss the implications of our results for understanding transient phenomena,
and in \S\ref{sect_conclusion} we summarize our conclusions.

\section{Input stellar evolutionary models and physical context}
\label{sect_input}

    \subsection{Input Models}

    The energy-deposition study presented here starts off from the stellar evolutionary calculations of WHW02, who
    provide a set of pre-SN massive stars, objects evolved all the way from the main sequence until the onset of core collapse.
We refer the reader to WHW02 for a discussion of the physics included in such models.
These physically-consistent models give envelope structures that obey the equations and principles of stellar
structure and evolution. The exact details of these models are not relevant since we aim at developing a general,
qualitative, understanding of stellar envelope behaviour in response to shock-heating - we do not aim at connecting a specific
observation to a specific input.
Here, we focus on 10-40\,\msun\, progenitor stars evolved at solar metallicity and in the absence of rotation. These calculations
include the effect of radiation-driven mass loss through a treatment of the outer-boundary mass-flux,
so that the final mass  in this set reaches a maximum of $\sim$15\,\msun\, for a
$\sim$20\,\msun\, progenitor star. Mass loss leads also to a considerable size reduction of the stellar envelope, with partial
or complete loss of the hydrogen rich layers, producing smaller objects at core collapse
for main-sequence masses above $\sim$30\,\msun. In the sequence of pre-SN models of WHW02,
surface radii vary from 4--10$\times 10^{13}$\,cm in the 10-30\,\msun\, range, decreasing to
$\sim 10^{11}$\,cm in the 30-40\,\msun\, range.

   This set of models is not exhaustive.
   All our input stellar-evolutionary models have reached the end of a massive-star life and all have a degenerate
   core that collapses. They are thus not exactly suited for post-main sequence massive stars that may be burning hydrogen/helium
   in the core/shell. The sampled mass range does not stretch to the most massive stars known, which also happen
   to exhibit variability and eruptions. In particular, it does not include massive blue supergiants,
   which are connected observationally with eruptive phenomena. It does not include highly-massive stars close to the Eddington
   limit, nor objects that for some reason would be near critical rotation. It does not include any object resulting from close binary
   evolution. These are clear limitations for an association of transient phenomena with specific progenitors, but this is not
   the goal of this study. Observations have likely been biased toward the most massive stars undergoing massive ejections,
   missing fainter progenitors undergoing more modest eruptions. Circumventing specific issues linking tightly specific models to specific
   observations, we attempt instead to develop a qualitative understanding. Our study is thus conceptual
   and explores the behavior of a gravitationally-bound envelope
   subject to shock-heating. As we explain below, the present set covers a wide enough range of properties to reveal important
   systematics in this context, which can then be applied more generally to other stars, be they of larger mass etc.

   \begin{figure*}
   \epsfig{file=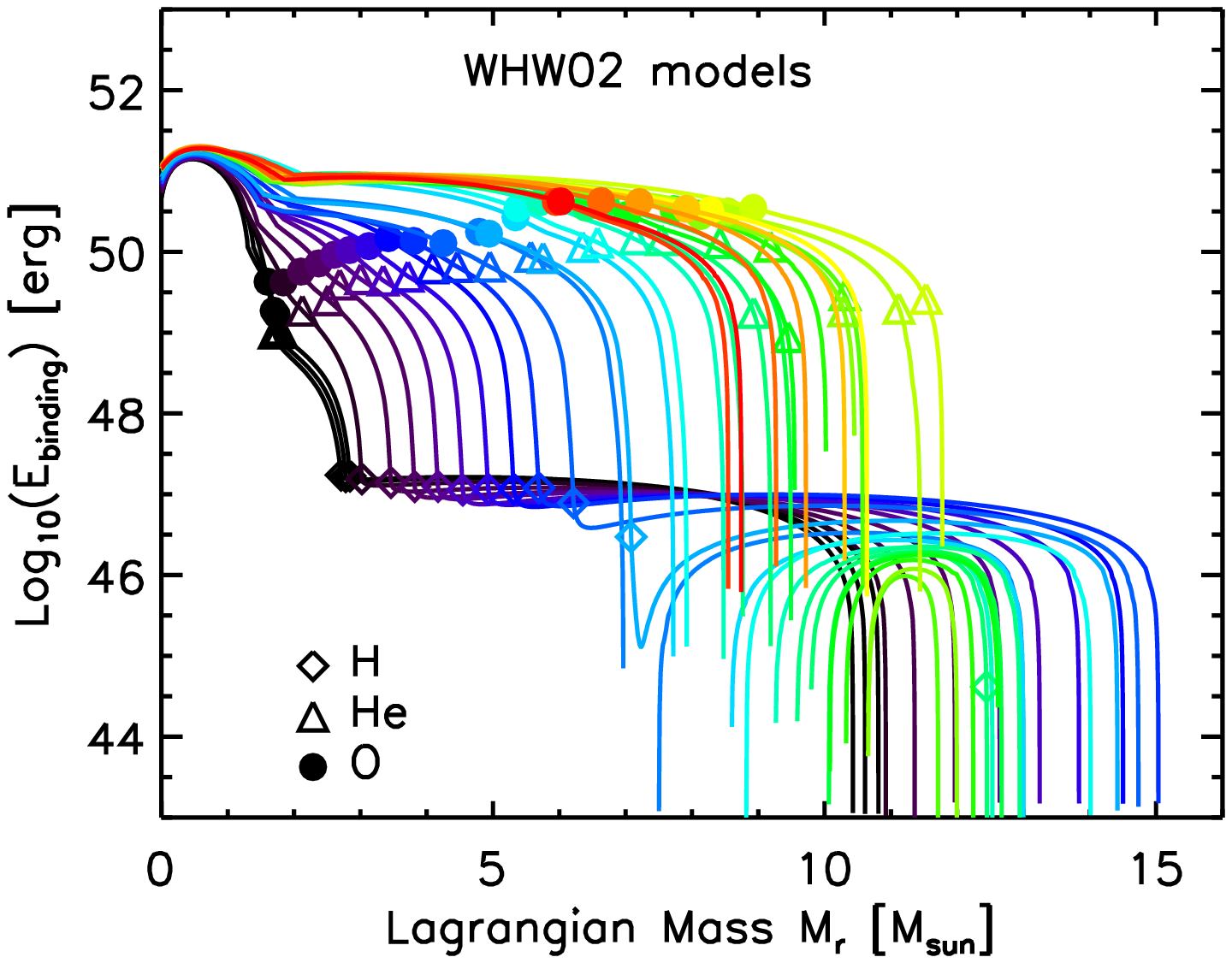,width=8.75cm}
   \epsfig{file=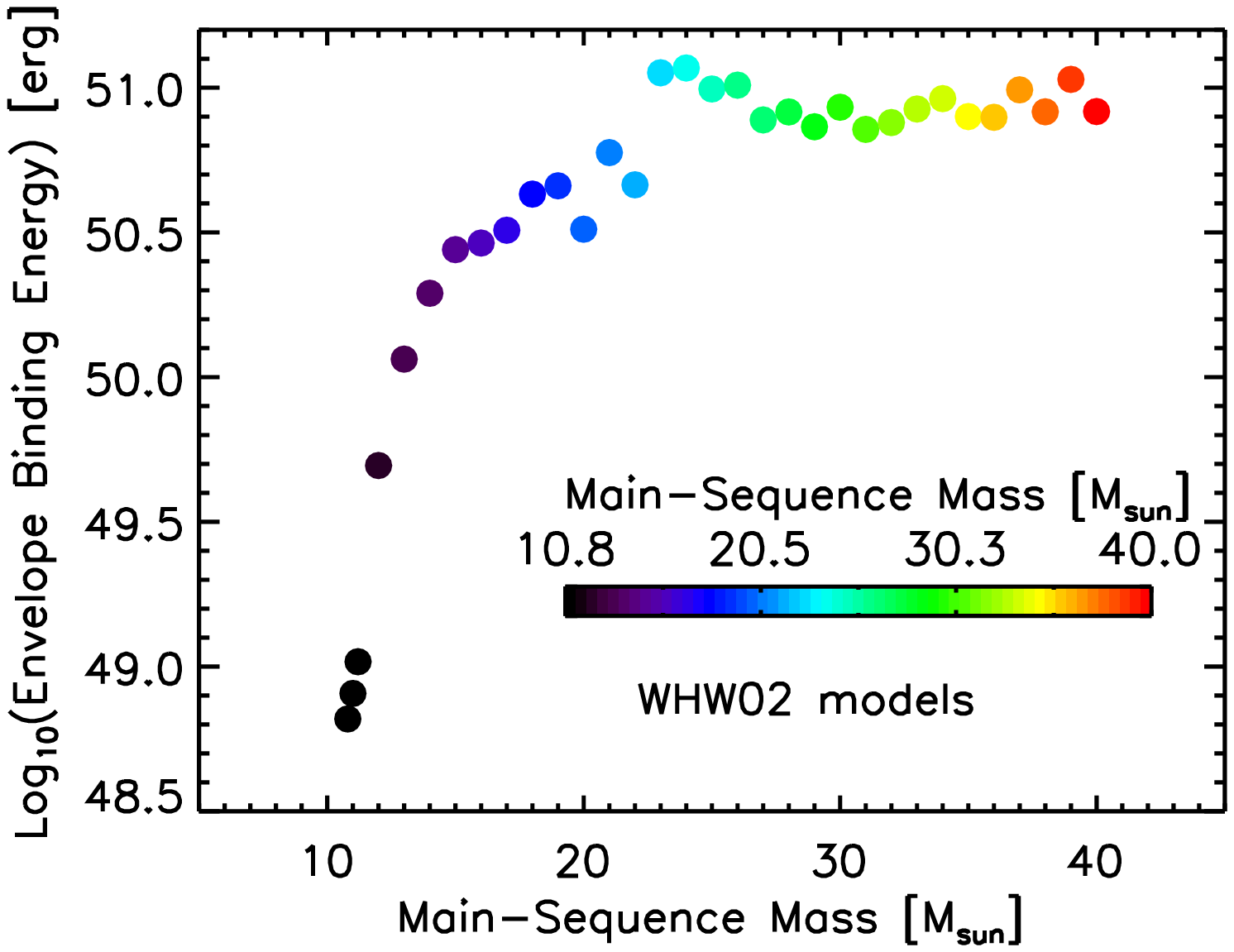,width=8.75cm}
   \caption{{\it Left:} Same as Fig.~\ref{fig_rho_mr}, but now for the binding energy as a
   function of the Lagrangian-mass coordinate $M_r$ of the envelope exterior to $M_r$:
   $e_{\rm binding}(M_r)  = \int_{r}^{r_{\rm max}} ( G M_r/r - e_{\rm int} ) dM_r$. In our determination
   of the binding energy, we have included the internal energy \citep{ZN_71}.
   Symbols refer to the base of the corresponding shells, with diamond for the
   H-rich shell, triangles for the He-rich shell, and dots for the O-rich shell.
   {\it Right:} Variation of the envelope binding energy at the onset of collapse in the models
   of WHW02 and shown as a function of the progenitor mass on the main sequence.
   Here, the ``envelope'' refers to total mass exterior to the inner 1.8\,\msun.
   The colorbar applies to both left and right panels.
   \label{fig_eb_mr}
   }
   \end{figure*}

    The density distribution above the iron core is an important property that distinguishes the structure of massive stars.
Low-mass massive stars show a steep density fall-off above their degenerate core, while the more
massive stars boast very flat density profiles (Fig.~\ref{fig_rho_mr}; recall that at the time shown,
the Fe or ONeMg core is degenerate and starts to collapse). In the context of core-collapse SN explosions, this has been
recognized as one of the key aspect of the problem, since this density profile directly connects to
the mass-accretion rate onto the proto-neutron star in the critical first second that follows
core bounce \citep{burrows_etal_07a}.
This has led \citet{burrows_goshy_93}, and more recently \citet{murphy_burrows_08} to identify
a criterion for explosion through a competition between neutrino luminosity and mass accretion rate.
Such a criterion, although tuned by the global hydrostatic configuration of the progenitor star, isolates the conditions
for a successful explosion to those in the direct vicinity of the pre-collapsed core. A more global, and more fundamentally
meaningful, criterion for a successful explosion is whether the energy deposited at the base of the progenitor envelope is smaller
or greater than its binding energy. This matter is rarely addressed explicitly, supposedly because the 10$^{51}$\,erg
explosion energy aimed for is thought to be much in excess of the binding energy of the envelope, but this may in fact
not be the case and it is ignoring the key role the binding energy plays in producing the diversity of characters observed
in core-collapse SNe, and potentially in numerous transients.

  \subsection{The Binding Energy Barrier}
\label{sect_ebind}

   The variation in the progenitor envelope density structure (Fig.~\ref{fig_rho_mr}) is echoed in the corresponding
   binding energy of the stellar envelope.
   In the left panel of Fig.~\ref{fig_eb_mr}, we show the binding energy  of the envelope exterior to a Lagrangian-mass coordinate $M_r$,
   as a function of $M_r$, for the same set of pre-SN massive star models.
  To differentiate mass shells of different compositions, we use symbols to mark the inner edge of the hydrogen shell (diamond),
   of the He-rich shell (triangle), and of the O-rich shell (dot).
   The binding energy shows a strong dependency with progenitor main-sequence mass, and we can identify two
   classes of objects. Objects with main-sequence mass below $\sim$25\,\msun\, possess a sizable hydrogen envelope,
   whose fraction of the total mass at the onset of collapse grows as we go to lower-mass progenitors.
   It represents at collapse $\sim$80\% of the total
   mass in the 11\,\msun\, model, but only $\sim$30\% in the 25\,\msun\, model (these masses refer
   to the main-sequence mass). In all these cases, whenever present, the binding energy of the hydrogen shell
   is very small, on the order of 10$^{47}$erg, hence very loosely bound to the star.
   In contrast, for objects with main-sequence masses in excess of 25-30\,\msun, the hydrogen envelope is shed through
   mass loss during the pre-SN evolution, and the pre-SN star is essentially a ``compact'' star, with an envelope whose
   binding energy is now 3-4 orders of magnitude larger. This is perhaps the single-most important characteristics
   of pre-SN massive star envelopes resulting from the single-star evolution calculations of WHW02,
   namely that hydrogen-deficient (generally higher-mass originally) stars are very
   tightly-bound objects while hydrogen-rich (generally lower-mass originally) stars are very loosely bound.

   Moving deeper into the envelope, i.e. in the He or O shells, the same trend persists, with a systematic increase
   in binding energy for each shell edge as we move up in mass from low-mass to high-mass massive stars.
   As shown in the right panel of Fig.~\ref{fig_eb_mr}, the binding energy of the entire envelope exterior
   to the degenerate core varies from $\lesssim$10$^{49}$\,erg for the 11\,\msun\, model (a RSG star with a final mass
   of  $\sim$10.6\,\msun) up to $\sim$10$^{51}$\,erg in the 40\,\msun\, (a H-deficient Wolf-Rayet star with
   a final mass of $\sim$8\,\msun) model.

   For an explosion to ensue, an energy greater than the binding energy must be deposited suddenly at the base of the envelope,
   the excess energy above the binding energy being used to accelerate the envelope and turn it into outflow/ejecta.
   While a very modest energy of  $\sim$10$^{49}$\,erg is sufficient to unbind the envelope of a 10\,\msun\, RSG star,
   such a deposition in a Wolf-Rayet star would produce nothing remarkable.
   In the present work, the connection between energy deposition and binding energy is key to the understanding
   of the properties of the resulting mass ejections. More generally, the low binding energy of massive star envelopes
   is of primary importance for the understanding of their stability/variability, as in the context of pulsations, for example.

     The origin of the energy deposition that we artificially treat in this work
    may be a sequel of the gravitational collapse of the core, in which case it can happen only once, leading either
    to an explosion or to a fizzle. Alternative sources in massive stars will likely be less energetic,
    but in objects with low binding energy, may still be of relevance
    to a wide spectrum of astrophysical events. The most suited mechanism in the present context would be the prompt
    thermonuclear combustion of a small amount of material.
    In the context of very massive stars like $\eta$ Car (not directly addressed through our progenitor set),
    \citet{guzik_05} propose that non-radial gravity mode oscillations could lead to mixing into the hydrogen-shell
    burning layer of fresh fuel located directly above it, thereby releasing suddenly a large amount of energy to
    unbind a fraction of the above layers. In lower mass massive stars at the end of their lives, this combustion could concern
    already processed material, like carbon, oxygen, or silicon,
    at a location just above the degenerate core. Such ``nuclear flashes'' have indeed been encountered  in stellar-evolutionary
    calculations of 8-12\,\msun\, massive stars \citep{weaver_woosley_79}.
   Interestingly, with a (nuclear) binding energy of $\sim$1\,MeV/nucleon, the combustion of $^{12}$C or $^{16}$O
   to $^{56}$Fe liberates $E_{\rm nuc} \sim1.9 \times 10^{49} (M / 0.01M_{\odot}$)\,erg. Hence, as little as a few percent
   of a solar mass of carbon or oxygen burnt to iron can yield an energy in excess of the binding energy of the lowest-mass massive stars,
   and of comparable magnitude to that of the weakest core-collapse SN explosions \citep{pastorello_etal_04,kitaura_etal_06}
   or LBVs.

    \subsection{Energy Transport by Diffusion/Convection versus Shock-heating}
\label{sect_tdiff}

\begin{figure}
\epsfig{file=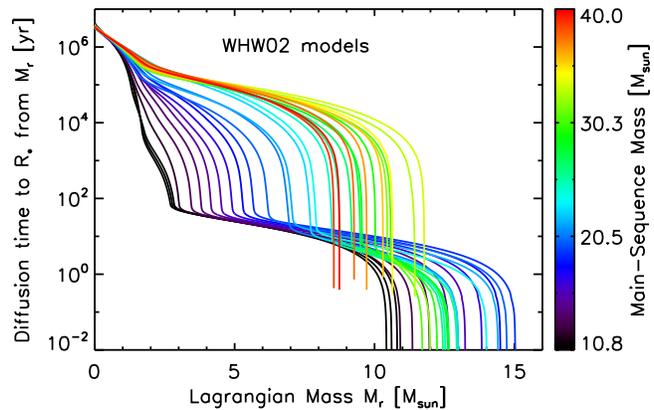,width=8.5cm}
\caption{Same as Fig.~\ref{fig_rho_mr}, but now showing the variation of the diffusion time.
The computation assumes unequal photon mean-free-path as in \citet{mitalas_sills_92},
but a constant and representative mass absorption coefficient of 0.1\,cm$^2$\,g$^{-1}$.
\label{fig_tdiff_mr}
}
\end{figure}

     In radiative stellar envelopes, energy is transported outward by diffusion. Any increase in energy release from
the core provokes a steepening of the temperature profile and an enhanced radiative flux, carrying outward this extra energy,
often aided by convection too. However, radiative diffusion or convection can only be effective for small increases
in energy, due to the short mean-free path of photons and/or the modest sound
speed. Associated time-scales for such energy transport
are therefore long and the means poorly efficient. In Fig.~\ref{fig_tdiff_mr}, we show the diffusion time as a function of depth
in the 10-40\,\msun\, model sequence of WHW02, computed by adopting a constant mass-absorption coefficient of 0.1\,cm$^2$\,g$^{-1}$
(intermediate between that for the hydrogen-rich and silicon-rich shells)
but a depth-dependent mean-free-path \citep[this is controlled primarily by the twenty-order-of-magnitude variation in density
between the core and the surface]{mitalas_sills_92}. Resulting diffusion times from regions immediately above the core
range from 10$^4$ to 10$^5$ years. In contrast, the shock crossing time-scale of the envelope is $\sim R_{\star}$/$\langle v_{\rm shock}\rangle$,
which for Wolf-Rayet to RSG progenitor stars with radii of 10$^{11}$ to 10$^{14}$\,cm and even modest, but supersonic,
shock waves with $\langle v_{\rm shock}\rangle \sim$1000\,\kms, is typically on the order of minutes to days.

\begin{figure}
\epsfig{file=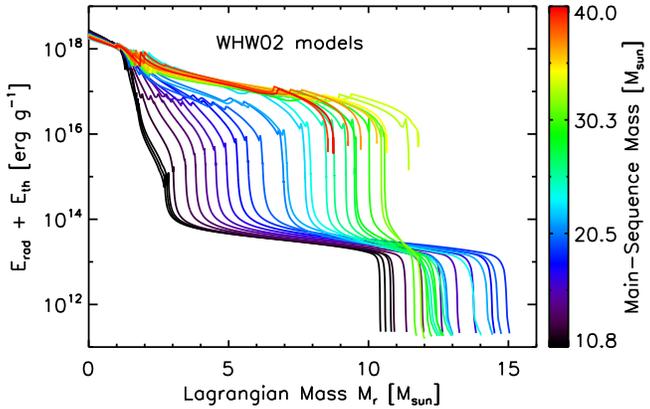,width=8.5cm}
\caption{Same as Fig.~\ref{fig_rho_mr}, but now showing the variation of the internal energy (including radiation and
thermal parts).
\label{fig_eint_mr}
}
\end{figure}

  In this study, we want  to explore what happens when the deposition of energy leads to an energy increase that cannot be
remedied either by diffusion or convection transport, but instead has to lead to shock formation (even for small energy deposition).
This leads to a completely different regime since in this situation the shock can communicate this extra energy to the {\it entire} envelope on
a short time-scale of days: The stellar envelope hence responds immediately to the change of conditions at depth, instead of secularly
evolving to a bigger/smaller or cooler/hotter star. This dichotomy was recently discussed in the context of the helium flash, whose associated energy
was found to be too small to lead to the formation of a shock \citep{mocak_etal_08,mocak_etal_09}.
In our artificial approach, the energy release has to occur on a very short
time-scale to trigger the formation of a shock.
How short will depend on the progenitor since, as shown in Fig.~\ref{fig_eint_mr}, the depth-dependent internal energy
(including both radiation and thermal components) varies
considerably between progenitors, increasing at a given $M_r$ with main-sequence mass. Forming a shock will require a stronger energy deposition,
a shorter deposition time, and/or a more exterior deposition site in higher mass progenitors.

\subsection{Origin of energy deposition}

In the present study on pre-SN massive-star progenitors,
  we find that an energy deposition that is marginally larger than the envelope binding
  energy yields ejecta that are reminiscent of SN impostors, but in the present context, are also reminiscent
  of the mass ejections that must take place prior to core collapse in interacting SNe. The short delay
  between the two ejections, of no more than a few years to produce a luminous interacting SN,
  suggests a shock-heating solution for both ejections. Any other energy-transport means in a massive star
  that could propel mass out of the gravitational potential would take place
on a time-scale of thousands
  of years or more. The disconnection between the surface and the core thus suggests
  that both events are tied to a phenomenon that happens in the core or in its direct vicinity.
  We propose thermonuclear flashes associated with shell burning in the last stages
  of massive star evolution (pair-instability pulsations are analogous to this scenario but may only
  apply to stars with a main-sequence mass in the range 95--130\,\msun; \citealt{woosley_etal_07}).
  These burning stages are very short (month to year time-scales) and always occur at the end of the life
  of all massive stars. They offer a natural tuning and a reproducible circumstance for the production of
  interacting SNe, with at least one major interaction between
  the penultimate shell ejection and that resulting from the gravitational collapse of the degenerate core.
  Given the relatively large number of interacting SNe compared to the strict time requirements
  to produce them  (a few-year delay is no more than an instant relative to the lifetime of a star!) suggests
  that indeed the conditions must be met often rather than encountered by chance.
  The stellar-evolutionary calculations of \citet{weaver_woosley_79} for 8-12\,\msun\, massive stars
  support the occurrence of nuclear flashes
  in association with Ne and Si core/shell burning, and occuring merely a few years prior to
  core collapse. These simulations are old and would need to be revisited in a thorough fashion,
  with full hydrodynamics, multi-dimensionality, and high resolution.

  In the context of more massive stars like $\eta$ Car, \citet{guzik_05} propose
  that non-radial gravity mode oscillations above the core and near the hydrogen-burning shell
  could lead to mixing of hydrogen-rich material downward into hotter denser layers. Combustion of
  this fresh fuel would yield a burst of energy triggering mass ejection and a bolometric brightening.
  In this work, we emphasize the dire consequences
  of any such energy release on loosely-bound massive stars, potentially triggering single eruptions
  as evidenced in transients or multiple eruptions as evidenced in interacting SNe \citep{dessart_etal_09}.

   More generally, the hydrodynamical investigations of the late stages of nuclear burning in massive
   stars by  \citet{bazan_arnett_94,bazan_arnett_98,asida_arnett_00,arnett_etal_05,meakin_arnett_06}
   reveal the multi-dimensional and time-dependent nature of the problem, issues that have been
   largely overlooked in stellar evolutionary calculations so far. In particular, they find that considerable
   energy can leak out of the convectively-burning shells in the form of waves, with a potential to alter
   the otherwise quasi-steady hydrostatic configuration of the stellar envelope. Similar investigations
   for lower-mass massive stars are highly needed.

   These works provide a physical motivation for our investigations. For the simulations presented
  here, we do not, however, build a fully-consistent picture of the problem, but instead
  start off by imposing the modalities of such an energy deposition. In practice, we follow a very simple
  approach, adopting a constant energy-deposition rate for a given duration. We thus
  neglect any feedback of the dynamics on the mechanism causing the energy release.

   \section{Numerical Technique and Computational Setup}
 \label{sect_model}

The numerical simulations presented in this work were all performed with
Vulcan/1D (V1D), a one dimensional radiation-hydrodynamics
code that works in the framework of Newtonian physics. It incorporates many methods
and techniques, using both Lagrangian and Eulerian schemes for
planar, cylindrical, and spherical geometries. The Lagrangian hydrodynamics
scheme is based on the staggered-grid method described in \citet{RM_67},
which uses artificial viscosity for shock waves. Some test problems
of an early version of this code are described in \citet{livne_93}.

For the treatment of non-LTE radiative-transfer problems we have
recently implemented several solvers with
different levels of approximations for the radiation fluxes. The coarser
method is our gray flux-limited-diffusion method, which has also a multi-group
extension. We have also constructed more accurate schemes for the
radiation fields using moment methods and full angle-dependent solvers,
which are similar in nature (but not in details) to the scheme described in \citet{ensman_phd,ensman_94}.
Those however were not implemented in this work. The radiation field
is coupled to matter in a fully implicit fashion, which guarantees stability
and large time steps.
Since the important physics relevant to this study occurs at large optical depth, the
multi-group capability of the code is not used.
Opacities/emissivities are interpolated from a set of tables prepared with a separate
program. It computes the populations for a large number of atomic levels (up to 1000 per species), under the assumption of LTE, for typically
fifteen species (H, He, C, N, O, Ne, Na, Mg, Si, S, Ar, Ca, Cr, Fe, and Ni), and for a set of compositions representative of the pre-SN
envelope of a 15\,\msun\, main-sequence star (we generate tables for compositions corresponding to a mean atomic weight between
1.3 and 25 atomic-mass units). For each species, up to 18 ionization stages are included.
At present, we account for scattering opacity (due to electrons), and absorptive opacity due
to bound-free and free-free processes. Although lines can be included, we have neglected them in this study.

The equation of state takes into account excitation/ionization to all available levels
of the different atomic/ionic species in the composition. The distribution function
of these levels is computed using a method similar to the
one described in \citet{kovetz_shaviv_94}. The resulting electron density is
then used together with the temperature in order to extract the pressure,
energy, chemical potential and their derivatives respective to the electron
density and temperature from a table computed in advance by solving the
Fermi-Dirac integrals. The pressure, energy and entropy of the ions are then
added as an ideal gas.

All our simulations use as inputs the massive-star models of WHW02, evolved from the main sequence until the onset of core collapse.
At present, V1D adopts the same grid as the one read-in from the chosen input model of WHW02. The resolution is therefore quite low
 at the progenitor surface, which affects the accuracy of the breakout signal. The range of mass encompassed by the V1D grid
 is set by the inner mass cut where energy is deposited, the excised region being shrunk to a point mass at the origin.
  We do not compute the explosive nucleosynthetic yields, and thus do not alter the original composition of the massive-star
 envelopes computed by WHW02. In particular, there is no production of $^{56}$Ni and no associated heating through
 radioactive decay accounted for in our simulations. In the main
 parameter study presented here, we use the 11\,\msun\, model (named s11.0 by WHW02). We first explore the evolution of
 the progenitor envelope after depositing an energy $E_{\rm dep}$ between 5$\times 10^{48}$\,erg and 1.28$\times 10^{51}$\,erg
 at a mass cut $M_{\rm cut}=1.8$\,\msun\, (uniformly deposited in the range 1.8 to 2.3\,\msun) and for a duration of
 10\,s (\S\ref{var_edep}). We also explore the sensitivity of the outcome to the location of the energy deposition, from a Lagrangian mass
 coordinate of 1.8 to 3, 7,  and 9\,\msun\, (\S\ref{var_mcut}), and to the duration of the energy deposition, from ten seconds, to one hour,
 one day, one week, and one month (\S\ref{var_dt}). Finally, we perform a few simulations using more massive progenitors,
 with 15, 20, and 25\,\msun\, main-sequence mass (\S\ref{var_mprog}). A summary of the model initial parameters as well as
 quantities characterizing the main results is shown are Tables~\ref{tab_s11} and \ref{tab_s11_wcomp}.
  Note that in the text, when we mention the masses associated with the WHW02 models, we mean the main-sequence masses,
 as in the 11\,\msun\, or the 25\,\msun\, models; these do not correspond to the star mass at the onset of collapse, which is shown
 for all models in, e.g., Fig.~\ref{fig_rho_mr}.

 \section{Results for the 11\,\msun\, sequence}
 \label{sect_s11}

\begin{table*}
  \begin{minipage}{160mm}
    \caption{Summary of the parameters and key results for the sequence started with the 11\,\msun\, progenitor model (pre-SN mass
    of 10.6\,\msun).
Note that all models are not run for the same total time ($t_{\rm end}$), owing to numerical problems at very late times as the density
and temperature drop in the outer ejecta. Numbers in parenthesis refer to powers of ten in the unit shown in the first column of the same row.
The effective mass on the V1D grid depends on the adopted inner mass cut $M_{\rm cut}$.
The time origin corresponds to the start of the simulation, when energy is deposited for $t_{\rm edep}$ seconds.
Starting from the top row down, we give the total energy deposited (in 10$^{50}$\,erg), the inner mass shell where it is deposited
and the duration of that deposition (in s), the time at the end of the simulation (in days),
the sum of the mass of all shells that at the end of the simulation have a velocity larger than the local escape speed $\sqrt{2 G M_r / r}$,
the ejecta mass-weighted average velocity (in \kms), the kinetic energy of the ejecta (in 10$^{50}$\,erg) at the end of the simulations,
the maximum ejecta velocity (in \kms), the time (in days) and the peak luminosity
(in \lsun) of shock breakout , the average speed of the shock (in \kms), the kinetic and the internal (thermal plus radiation contributions)
at the time of breakout (in erg), the time of peak luminosity in the post-breakout plateau (in days) and the corresponding luminosity (in \lsun),
the duration of high brightness for the transient (time during which the luminosity is greater than1/50th of $L_{\rm peak, plateau}$),
and the time-integrated bolometric luminosity (in 10$^{49}$\,erg; note that the time interval varies
between models). [See text for discussion.]
\label{tab_s11}}
\begin{tabular}{lcccccccccc}
\hline
Model Name                    & s11\_0  & s11\_01   & s11\_1   & s11\_2    & s11\_3   & s11\_4    & s11\_5    & s11\_6 & s11\_7  & s11\_8  \\
\hline
$E_{\rm dep}$  (10$^{50}$\,erg)         & 5.0(-2)   & 7.5(-2) &   1.0(-1) & 2.0(-1)   & 4.0(-1)   & 8.0(-1)    & 1.6(0)  & 3.2(0)  & 6.4(0)& 1.28(1)\\
$M_{\rm cut}$  (\msun)      & 1.8    & 1.8    & 1.8    & 1.8    &  1.8   & 1.8     & 1.8     & 1.8   &  1.8 & 1.8    \\
$t_{\rm edep}$     (s)               & 10   & 10   & 10   & 10   & 10   & 10   & 10   & 10  & 10  & 10   \\
$t_{\rm end}$ (d)               &  730   & 193      & 730     & 730     & 160     & 730      & 117    & 103     &  730   & 686  \\
$M_{\rm ejected}$  (\msun)      & 0.13    & 5.95    & 7.75    & 8.60     &  8.72   & 8.79     & 8.80     & 8.80   &  8.80 & 8.80    \\
$\langle v \rangle_{M}$  (\kms)            &  42      &  76   & 121     & 296      & 510     & 789      & 1165    & 1687   & 2412  & 3432  \\
$(E_{\rm kin})_{\rm end}$  (10$^{50}$\,erg)  & 2.30(-5)&4.00(-3) & 1.33(-2) & 9.25(-2) & 2.79(-1) & 6.74(-1) & 1.47(0)&3.08(0) & 6.30(0) & 1.276(1) \\
$\langle v_{\rm max}\rangle$  (\kms) & 60 & 410  & 560 & 1310    & 1800   & 2740    & 4100   & 5970  & 8740 & 12700  \\
$t_{\rm SBO}$ (d)               & 54.7     & 32.1   & 21.4    & 9.6     & 5.8    & 3.8     & 2.6     & 1.8    & 1.3   &  0.9    \\
$L_{\rm peak, SBO}$ (\lsun)     & 1.7(5)  & 3.3(6)  & 5.2(7)  & 1.2(9) & 4.9(9)  & 1.6(10) & 4.2(10) &1.1(11)& 2.5(11)& 5.7(11)  \\
$\langle v_{\rm shock}\rangle$  (\kms) & 86 & 147  & 221     & 492    & 812    & 1238    & 1815  & 2622 & 3717 & 5244 \\
$(E_{\rm kin})_{\rm SBO}$ (erg) & 2.1(46) & 2.1(47) & 6.7(47) & 4.4(48) & 1.3(49) &  3.1(49) &  6.7(49) &  1.4(50) & 2.8(50) & 5.8(50) \\
$(E_{\rm int})_{\rm SBO}$ (erg)  & 5.2(48) & 4.6(48) & 3.8(48)  & 7.5(48) & 1.7(49) & 3.9(49) & 8.3(49)  &  1.7(50) & 3.5(50) & 7.1(50) \\
$t_{\rm peak,plateau}$ (d)      & 208      & 198     & 243     & 115     & 97      & 82       & 73       & 64    &  55     & 49     \\
$L_{\rm peak, plateau}$ (\lsun)  & 3.6(5)  & 1.5(6)  & 2.9(6) & 1.2(7) & 2.9(7)  & 6.0(7)  & 1.1(8)   & 2.0(8)&  3.6(8) & 6.2(8) \\
$\Delta t_{\rm plateau}$ (d)      &   675  & 517  & 341 & 287 & $>$155  & 180 & $>$115  & $>$106 & 111 & 100     \\
$\int L_{\rm bol}$ dt (10$^{49}$\,erg) & 6.6(-3) & 5.6(-3) & 2.5(-2)& 6.7(-2)  & 1.3(-1)  & 2.5(-1)  & 4.0(-1)   & 6.6(-1)& 1.1(0) & 1.8(0) \\
\hline
    \end{tabular}
  \end{minipage}
\end{table*}

  \subsection{Effect of varying the Energy Deposition Magnitude}
\label{var_edep}

  In this section, we present the results for the sequence of simulations based on the 11\,\msun\, model for a range of energy
  deposition magnitudes. A compendium of parameters describing the set up and
  the results is given in Table~\ref{tab_s11}. Note that, initially (prior to energy deposition),
  the envelope (total) internal energy is $\sim 10^{49}$\,erg, and its gravitational binding energy is
  -2.2$\times 10^{49}$\,erg in this 11\,\msun\, model (the model mass at the time of collapse is 10.6\,\msun).

    In our simulations, energy deposition leads to a strong increase in internal energy (temperature, pressure),
    leading to the formation of a shock which moves outward.
    In all runs, the shock crosses the entire envelope and eventually emerges at the surface of the progenitor, giving rise
    to a shock-breakout signal (more precisely, a radiative precursor precedes the emergence of the shock, but one generally
    refers to this whole phase as shock breakout).
    The shocked envelope left behind has been turned into a radiation-dominated
    plasma, with an extra energy that varies from a small ($E_{\rm dep}=5\times 10^{48}$\,erg in the s11\_0 model)
    to a large value ($E_{\rm dep}=1.28\times 10^{51}$\,erg in the s11\_8 model).

\begin{figure}
\epsfig{file=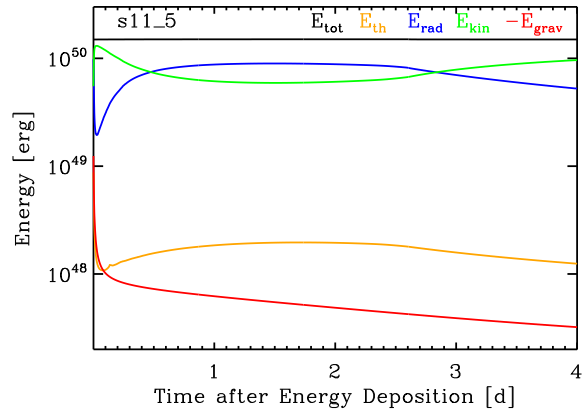,width=8.5cm}
\caption{Time evolution of the mass-integrated total (black), thermal (orange), radiative (blue), kinetic (green), and
gravitational (red; its absolute value) energies for the model s11\_5. Notice the interplay between radiative (which dominates the
internal energy over the thermal component) and kinetic energy. The time of shock breakout is 2.6\,d,
at which the total internal (kinetic) energy is 8.3$\times 10^{49}$\,erg (6.7$\times 10^{49}$\,erg).
The time origin is the start of the simulation.
\label{fig_s11_5_enr}
}
\end{figure}

   After energy deposition, all the stored energy is internal. As the shock forms and moves out, that stored internal
   energy is converted into kinetic energy behind the shock, and as the shock progresses outward, more and more material
   is shock-heated and the total internal energy goes up at the expense of the kinetic energy.
   As time progresses further, the kinetic energy eventually increases due to radiation work on the ejecta.
   Figure~\ref{fig_s11_5_enr} illustrates this evolution until just after shock breakout for the case in which
   $E_{\rm dep}=1.6\times 10^{50}$\,erg (model s11\_5).

   As the shock approaches the surface, the envelope properties become qualitatively and quantitatively different,
   depending on the adopted $E_{\rm dep}$ value. For a large $E_{\rm dep}$,  we find that at shock emergence,
   the total envelope energy is in rough equipartition between kinetic and internal. As time progresses, all the energy
   is eventually converted into kinetic as radiation, nearly entirely trapped  at large optical depth, does work to
   accelerate the ejecta to its asymptotic velocity.

   In contrast, for small $E_{\rm dep}$, the kinetic energy of the envelope at breakout and later is very small. Here, the stored internal
   energy is weakly enhanced after energy deposition (i.e. $E_{\rm dep} \lesssim E_{\rm int}$ originally),
   and this excess energy is essentially all exhausted in doing work against gravity,
   merely lifting off the envelope from the gravitational potential well. At shock emergence, the envelope kinetic energy
   is thus negligible.

   Hence, while in all cases from low to high energy deposition, a shock forms and eventually emerges at the progenitor
   surface, the quantitative differences are very large between cases. Transitioning from dynamic to quasi-static diffusion regimes
   of energy transport, we identify three situations:
   1) The energy deposition is much larger than the binding energy and a SN-like explosion ensues; 2) the energy
   deposition is on the order of the binding energy and a low-energy explosion/eruption results; and 3) the
   energy deposition is lower than the binding energy and the excess energy merely shifts the star to a new quasi-equilibrium
   state from which it relaxes on a very long time-scale. In the following subsections, we describe each of these regimes individually.

   \subsubsection{$E_{\rm dep} > E_{\rm binding}$}

    This regime applies to our simulations with $E_{\rm dep}$ between 4$\times 10^{49}$\,erg (model s11\_3)
    and  1.28$\times 10^{51}$\,erg (model s11\_8). The separation between this sample and cases with lower
    energy deposition is drawn at the threshold value of $E_{\rm dep}$ for which the entire envelope is ejected.
    In Table~\ref{tab_s11}, we give a census of the properties for this set. Although qualitatively similar, these
    simulations give rise to 8.8\,\msun\, ejecta with a mass-weighted average velocity in the range 500--3500\,\kms,
    and kinetic energy in the range 2.8$\times 10^{49}$ ($E_{\rm dep}=4\times 10^{49}$\,erg) up to
    1.276$\times 10^{51}$\,erg ($E_{\rm dep}=1.28\times 10^{51}$\,erg). For the former, the binding energy
    takes a visible share of the energy deposited so that the asymptotic kinetic energy is sizably smaller than $E_{\rm dep}$.
    For the latter, the energy deposited is overwhelmingly large and represents 99.7\% of the final kinetic energy.
    In all cases, the initial prompt deposition of energy turns the stellar envelope into a true ``ejecta'',
    and this material is lost from the star without any noticeable fallback.

    We find that the shock-crossing time through the envelope is between
    0.9 and 5.8\,d (the distance between the deposition site and the progenitor surface is 4.7$\times 10^{13}$\,cm),
    with a time-averaged shock velocity between $\sim$5000 and 800\,\kms (in the same order
    from high to low $E_{\rm dep}$).\footnote{Note that the shock is always supersonic and may propagate
    at very different speeds depending on location and initial energy deposition.}
    This therefore modulates the delay between the energy release (which
    may be a sequel of gravitational collapse or otherwise) and the time of shock breakout, which is the first detectable signal
    testifying for the violent energy deposition that took place at depth. Importantly, this breakout signal is
    the unambiguous signature that a shock wave emerged from the progenitor surface, a signature that would
    directly eliminate steady-state wind solutions for the origin of the subsequent outflow/ejecta.

\begin{figure}
\epsfig{file=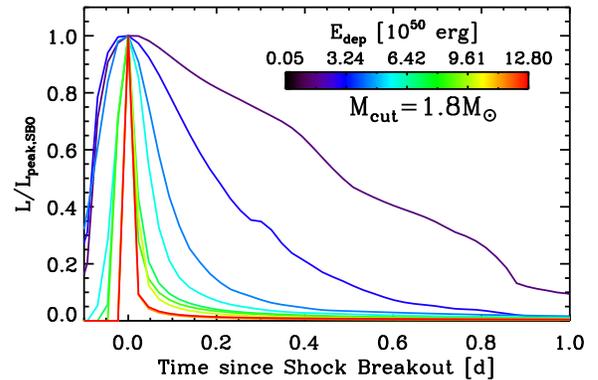,width=8.5cm}
\caption{Time evolution of the normalized intrinsic bolometric luminosity around shock breakout, as computed with
V1D, and revealing the hour-to-day range in duration of the shock-breakout signal, caused here
by the change in shock speed (or explosion energy). Light-travel-time delays are not accounted for.
The variation in the breakout-peak luminosity is shown in Fig.~\ref{fig_s11_lum_all}.
A color coding is used to differentiate models.
\label{fig_s11_lum_sbo}
}
\end{figure}

\begin{figure*}
 \epsfig{file=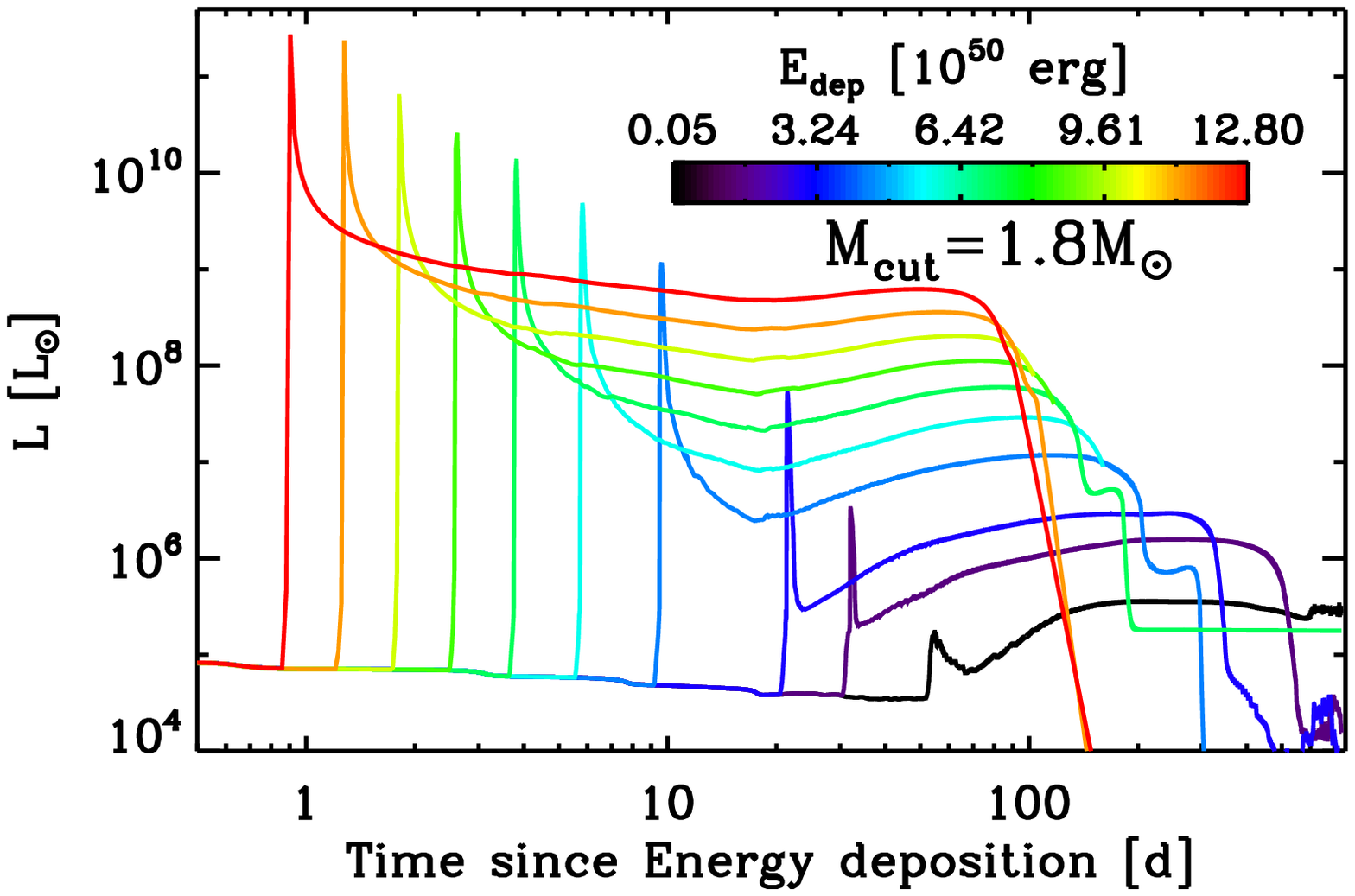,width=8.5cm}
 \epsfig{file=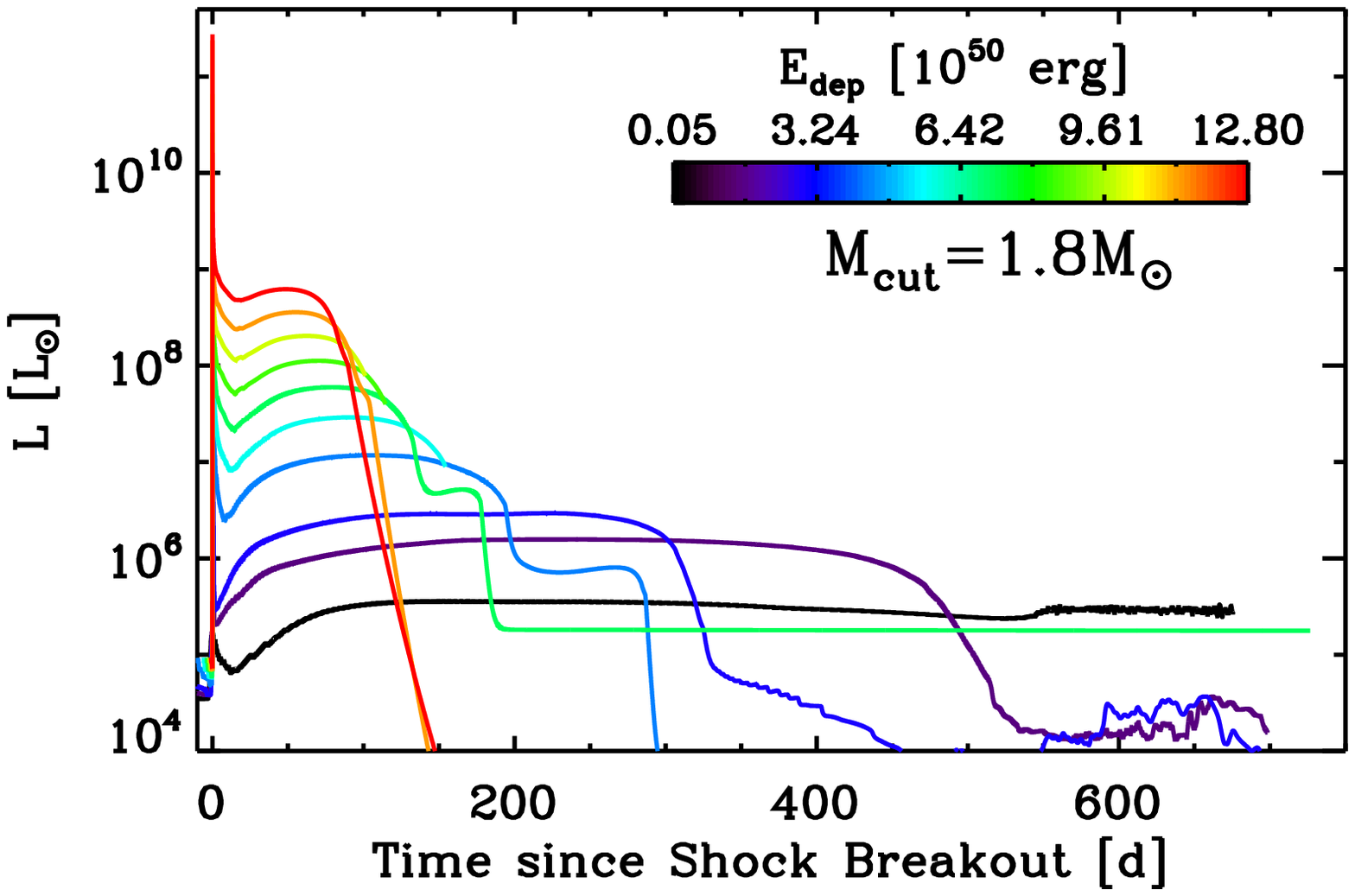,width=8.5cm}
 \caption{Same as Fig.~\ref{fig_s11_lum_sbo}, but this time showing the un-normalized bolometric luminosity.
 In the left panel, we use a logarithmic scale for the abscissa, showing the time since the start of the simulations.
 In the right panel, we use a linear scale for the abscissa, showing the time since shock breakout.
 Stronger explosions show shorter and stronger breakout signals, followed by a brighter plateau brightness
 but shorter plateau length. At intermediate values of the energy deposition, the plateau-brightness is followed
 by a ledge, which corresponds to the time when the photosphere recedes into the more tightly-bound
 (and slowly-expanding) helium-rich part of the envelope - no artificial nor physical mixing
 occurs in our 1D simulations, a process that could smear out such a feature.
 Note that for cases in which $E_{\rm dep} \sim E_{\rm binding}$, the breakout luminosity
 is barely larger than the following plateau, which we find here can last for up to two years in this
 11\,\msun\, progenitor star.
 \label{fig_s11_lum_all}
 }
 \end{figure*}

   Owing to the sizable range in shock speeds, the breakout signal varies considerably in duration. From a random walk
   argument, it is born when the decreasing envelope optical-depth $\tau$ at the shock
   depth $\Delta r$ below the surface eventually makes the diffusion time $t_{\rm diff} = \tau \Delta r /c$ shorter than the shock-propagation time to the
   progenitor surface $t_{\rm shock} = \Delta r / v_{\rm shock}$. This occurs at an optical depth $\tau \sim c / v_{\rm shock}$, and can be very large
   for a small shock-propagation speed. As shown in Fig.~\ref{fig_s11_lum_sbo} (note that the luminosity in this figure is normalized to the peak value
   to better illustrate its time variation), we find shock-breakout durations from $\sim$0.5 up to $\sim$2\,hr in
   the sequence (we refer to the intrinsic breakout duration, hence neglect light-travel time effects which would broaden the signal
   seen by a distant observer over a duration of at least $\sim R_{\star}/c \sim$1\,hr).
   Similarly, the peak luminosity at breakout varies enormously with the energy deposited or the shock strength, as shown in
   the left panel of Fig.~\ref{fig_s11_lum_all}, where the time axis is the time since energy deposition.
   To a higher energy-deposition strength corresponds stronger shocks,
   shorter shock-crossing times (earlier shock emergence), and both shorter-duration and
   greater peak breakout luminosities. For the range of models discussed
   in this section, we obtain values between 4.9$\times 10^{9}$ up to 5.7$\times 10^{11}$\,\lsun, a contrast of about a hundred which is comparable
   to the corresponding contrast in asymptotic ejecta kinetic energy. So far, the systematics of shock breakout
   have been tied to variations in the progenitor properties such as surface radius or atmospheric scale height, while obviously the explosion energy,
   if stretched to low enough values, can lengthen considerably the breakout signal.
   But importantly, while the delay between energy deposition and breakout varies by a factor
   of a hundred in this model set, it remains on a time-scale of days and hence much shorter
   than the diffusion time on the order of thousands of years for the corresponding layers (see \S\ref{sect_tdiff}).

\begin{figure*}
\epsfig{file=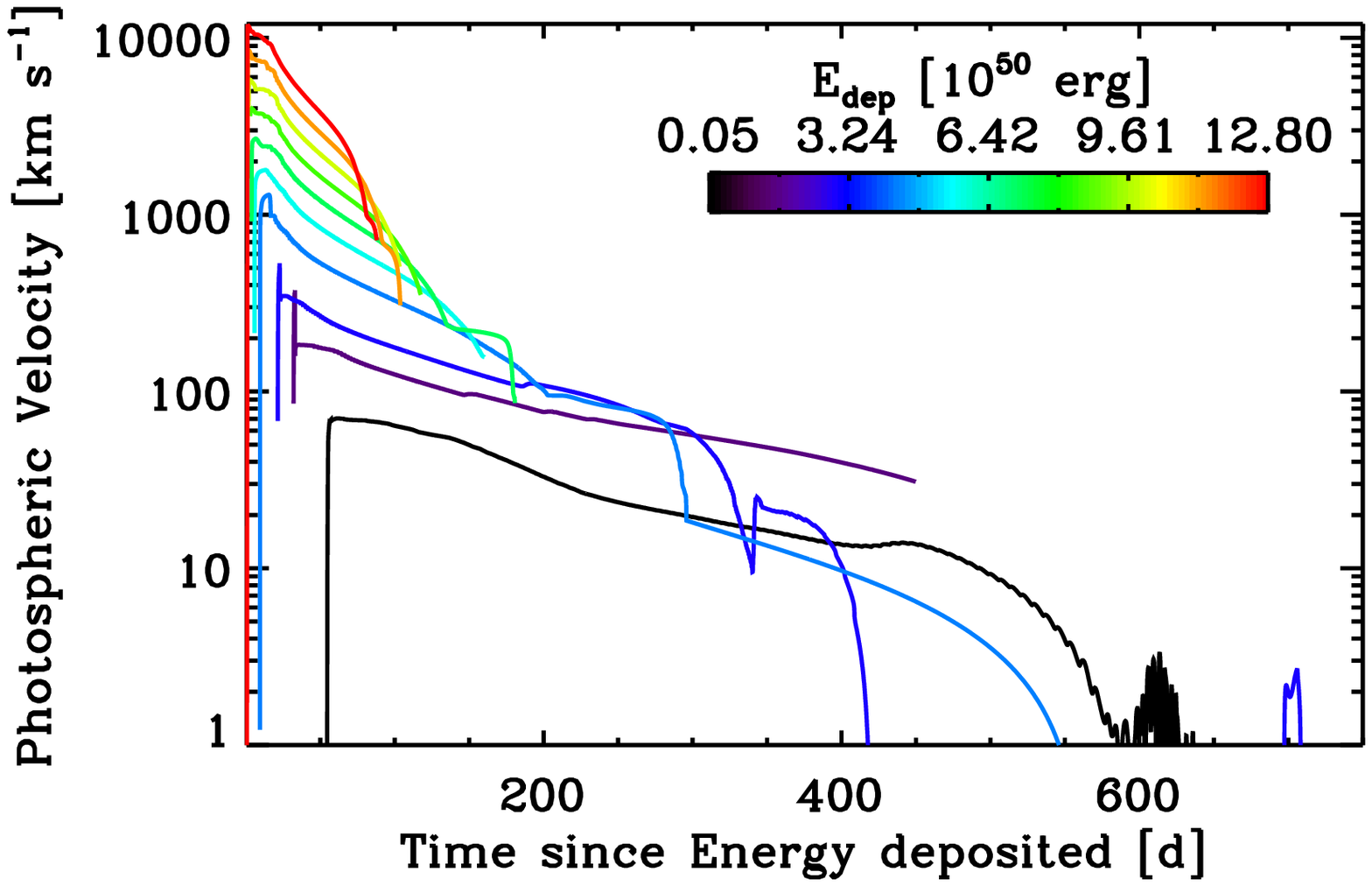,width=8.5cm}
\epsfig{file=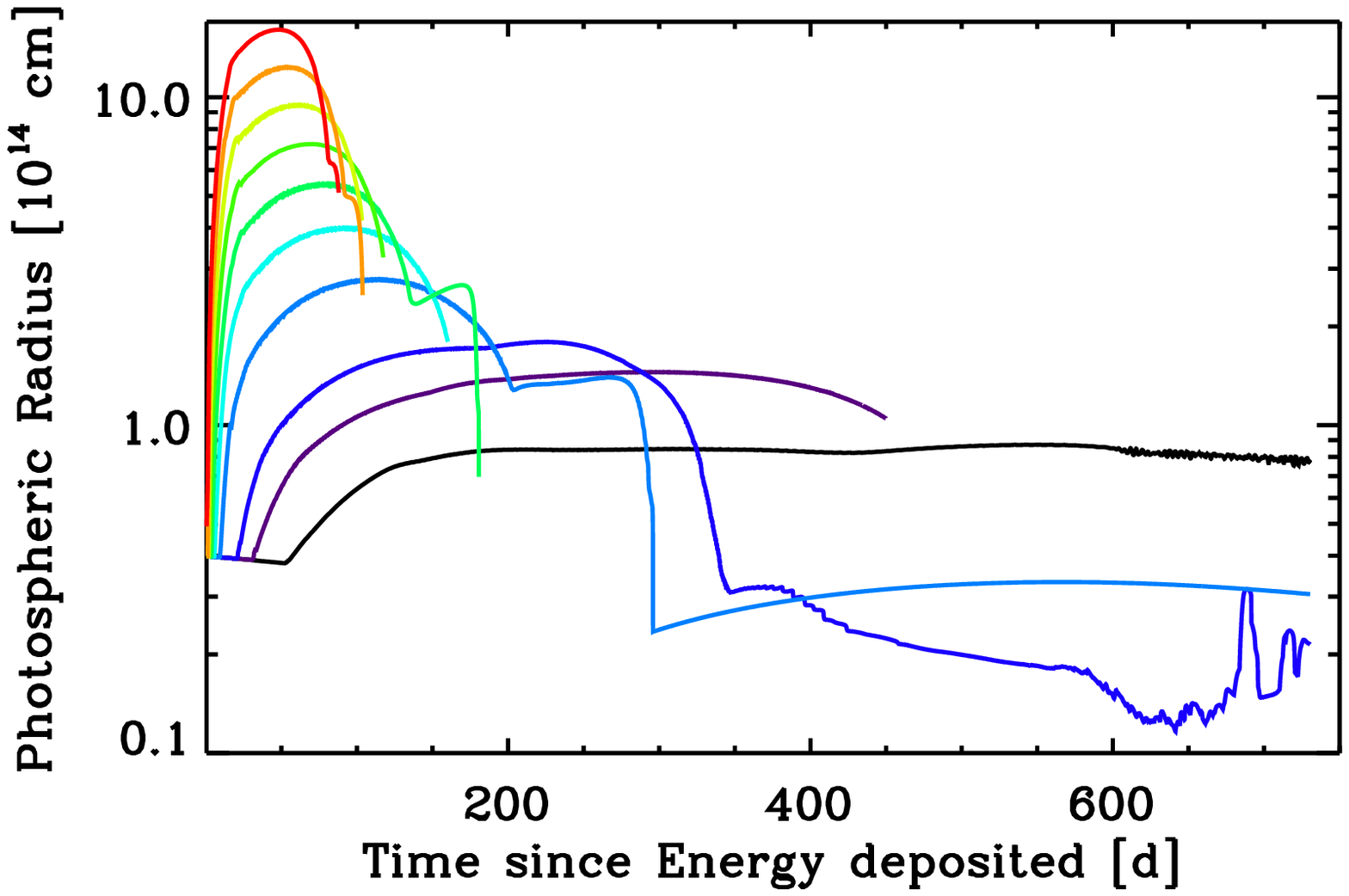,width=8.5cm}
\epsfig{file=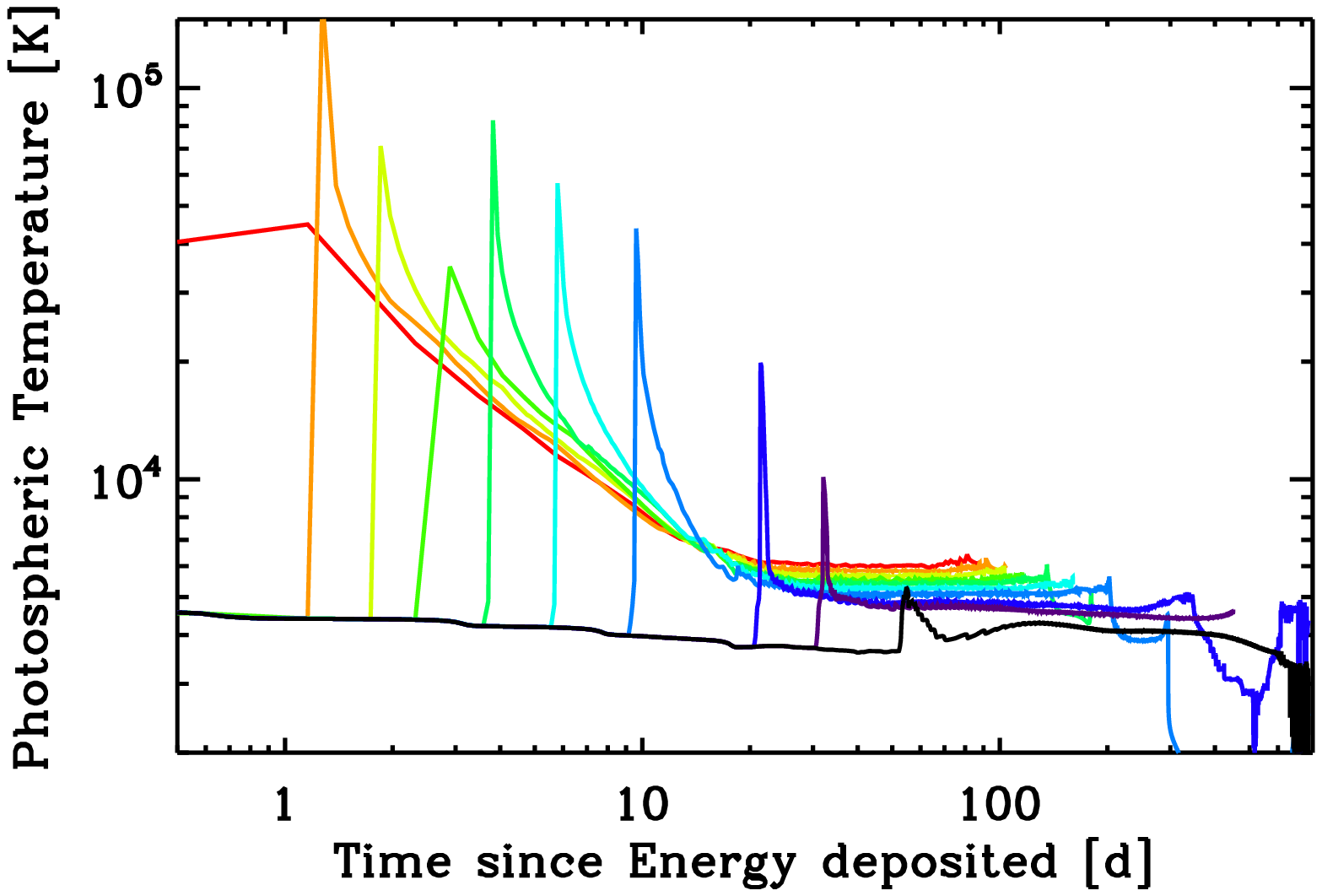,width=8.5cm}
\epsfig{file=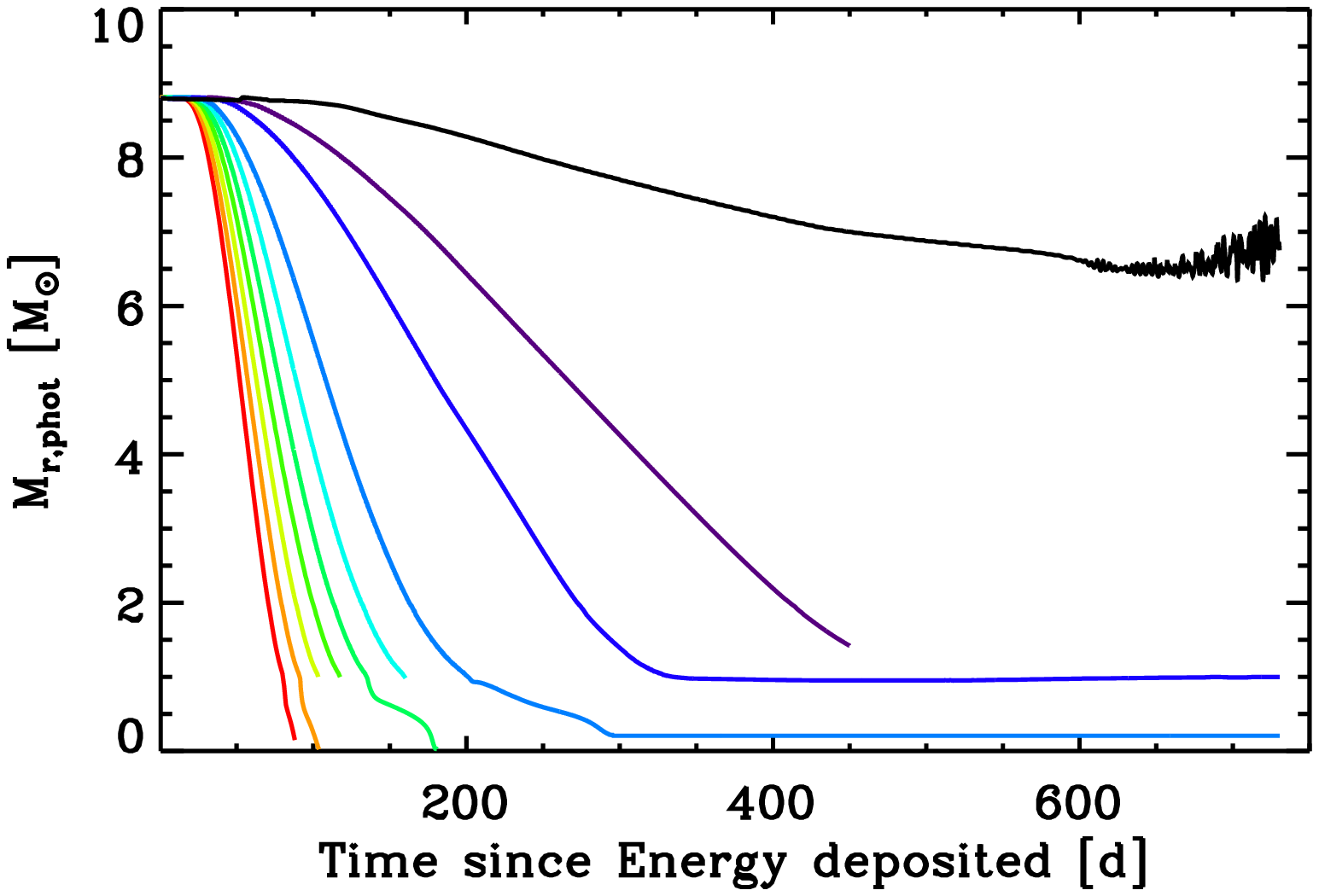,width=8.5cm}
\caption{{\it Top left:} Time evolution of the velocity at the photosphere for the sequence of
models associated with the 11\,\msun\, model of WHW02. Only the energy deposited
during the initial ten seconds after the start of the simulation differs between models,
all initial model characteristics being the same. A color coding is used to distinguish
the different cases, with total energy deposition encompassing the range 5$\times 10^{48}$
(black) up to 1.28$\times 10^{51}$\,erg (red).
{\it Top right:} Same as top left, but now for the radius at the photosphere.
{\it Bottom left:} Same as top left, but now for the temperature at the photosphere.
Note that the time of breakout is poorly sampled (dumps written every 10000\,s), so that
the temperature shown at breakout is an underestimate.
{\it Bottom right:} Same as top left, but now showing the Lagrangian mass coordinate
of the photosphere as a function of time. Note that the mass interior to the inner boundary of the
V1D grid is 1.8\,\msun\, for this sequence of models, and is not accounted for in this last panel.
\label{fig_s11_phot}
}
\end{figure*}

    These properties at and subsequent to breakout are visible in the conditions at the photosphere, which we show
    in Fig.~\ref{fig_s11_phot} for the velocity $V_{\rm phot}$, the radius $R_{\rm phot}$, the temperature $T_{\rm phot}$,
    and the Lagrangian mass coordinate $M_{\rm phot}$. Because of the infrequent
    writing of model quantities to file, the breakout phase is not well resolved. This matters little for the velocity,
    the radius, and the mass at the photosphere, which all evolve slowly. However, the photospheric
    temperature shows a peak whose strength is underestimated in the high energy-deposition cases, characterized
    by short breakout durations (note that the low spatial resolution at the progenitor surface prevents an accurate modeling
    of this breakout phase in any case). Hence, at breakout, we find an approximate range of photospheric temperatures
    between 5$\times 10^4$ to multiples of 10$^5$\,K. In scattering-dominated atmospheres,
    the color temperature of the emergent spectral-energy distribution corresponds roughly to the gas temperature
    at the thermalization depth \citep[p 149]{mihalas_78}, which is up to a factor of two greater
    than $T_{\rm phot}$ \citep{DH05_epm}. For a thermal (blackbody) emitter at $T$, the peak of the spectral-energy
    distribution (SED) is at $\sim 2900/T_{4}$\,\AA\, (where $T_4$ is $T/10^4$ K).
    At breakout, the spectral-energy distribution for the present set of models will thus peak in the range 100-600\AA.
    It is at breakout that a fraction of the material at the progenitor surface gets accelerated to large velocities.
    In the sequence of models s11\_3 to s11\_8, we find increasing maximum ejecta
    speeds from 1800\,\kms\, up to $\sim$13000\,\kms, typically a factor of four higher than the mean mass-weighted velocity
    of the ejecta.

    The general evolution of the long-term, post-breakout, photospheric conditions is qualitatively similar for this set of models.
    After breakout and for a few days, the internal energy stored in the shock-heated stellar envelope does work to
    accelerate the ejecta to its asymptotic velocity. For larger energy deposition, the expansion rate is greater
    and leads to more efficient  cooling of the ejecta, thereby mitigating the impact of the larger internal energy initially provided by the shock.
    This cooling is quasi-adiabatic since little radiation is lost from the photon decoupling layers - it all occurs at large optical depth.
    After a few days, the ejecta velocity increases monotonically with radius (homologous expansion is only
    approximately attained since the progenitor radius may represent up to 5--10\% of the current photospheric radius at all times).
    As the ejecta expand, cool, and recombine to a lower ionization state, the photospheric velocity decreases with time.
    The photospheric radius first increases, reaching values in excess of 10$^{15}$\,cm
    for the high energy cases, but eventually decreases due to recombination and the diminishing optical depth of the ejecta.
    The location of the photosphere is that of full ionization, which for the H-rich composition of this stellar envelope occurs
    around 5000\,K. Hence, the photospheric temperature, after its initial peak at breakout, slowly decreases to level off
    at $\sim$5000\,K. Ultimately, the photosphere reaches the inner region of the ejecta and the ``event''
    enters its nebular phase. This transition occurs earlier for higher values of $E_{\rm dep}$,
    after about 100 days in model s11\_8 but up to $\sim$200 days in the s11\_3 model (since the code crashed for that model,
    this is an estimate based on the noticeable trend with energy deposition for that quantity; see Table~\ref{tab_s11}).

    This evolution in photospheric properties parallels the bolometric evolution of the explosion, which we show in
    Fig.~\ref{fig_s11_lum_all}. In all cases, the early breakout peak in luminosity is followed by a fading (initially from radiative
    cooling at the shock-heated surface layers) followed by a sustained brightness as the ejecta expand (the rate of increase
    in photospheric radius compensates the rate of decrease in photospheric temperature): This is the so-called plateau
    phase of Type II-P SNe. Peak luminosities, reached at a later time for a lower energy deposition (from 50 to 100 days
    after breakout), are in the range 3$\times 10^7$ up to 6$\times 10^8$\,\lsun.
    Similarly, the plateau duration lengthens with lower energy deposition, ranging from 100 up to about 200 days
    (recall that this occurs in models s11\_3 to s11\_8 for the same ejected mass of $\sim$8.8\,\msun).
    At the end of our simulations, we find time-integrated bolometric luminosities in the range 10$^{48}$--10$^{49}$\,erg,
    which represent, in the same order, 1/20th down to 1/70th of the corresponding asymptotic ejecta kinetic energy.

    Hence, at the high energy end, we obtain light curves that are reminiscent of Type II-P SNe (Fig.~\ref{fig_obs_lc}),
    ranging from the energetic events like SN 2006bp \citep{dessart_etal_08}, to the more standard
    SN 1999em \citep{leonard_etal_02a}, through to the low luminosity SN 1999br \citep{pastorello_etal_09}.
    The luminosity contrast of a factor of $\sim$20 in this set of models is on the order
    of that inferred for Type II-P SNe (Fig.~\ref{fig_obs_lc}; \citealt{pastorello_etal_09}). For moderate to high $E_{\rm dep}$ values,
    the plateau durations we
    obtain are on the order of those observed for Type II-P SNe, although we find a lengthening for lower $E_{\rm dep}$ values that has
    not been observed definitely. For example, the observations during the plateau phase of SN1999br are
    unfortunately truncated after 100 days, at a time when the SN was not showing a clear sign of fading.
    The subsequent observation 250 days later, when the SN has faded, does not exclude 99br kept its plateau
    brightness for a longer time, perhaps for a total of 150 or 200 days (see Fig.~\ref{fig_obs_lc}).

    To summarize, the regime of explosions described in this section corresponds to that of type II-P SNe,
with a range of explosion energies and bolometric luminosities that are in line with inferences from observations.
 The rather large energy-deposition magnitude (i.e. $\geq 4 \times 10^{49}$\,erg) in this regime
favors a scenario involving the gravitational collapse of the stellar core
and, given the resulting complete envelope ejection, the formation of a neutron star remnant.
We wish to emphasize that there is no fundamental limitation
for preventing explosions in this 11\,\msun\, model for energies as low as a few times 10$^{49}$\,erg, as in model s11\_3.
The corresponding light curve would be reminiscent of that of SN 1999br, only slightly fainter.

   \subsubsection{$E_{\rm dep} \sim E_{\rm binding}$}

   In this section, we continue the sequence toward lower energy deposition, entering a regime where
   its magnitude is on the order of the binding energy of the envelope, and, incidently, on the same order
   as the total internal energy in the envelope. In contrast with results for high $E_{\rm dep}$ values, only
   a fraction of the progenitor envelope is ejected in models s11\_01 to s11\_2, ranging from 5.95 up to
   8.6\,\msun. Despite the incomplete success for explosion, all models go through a similar sequence
   as that of powerful explosions. A shock crosses the progenitor envelope, heating it and propelling
   it outward. The shock eventually breaks out, after 10--30\,d (time-average shock speeds of 150--500\,\kms),
   with an accompanying peak luminosity of 3$\times 10^6$ up to 10$^9$\,\lsun. The fastest ejected material
   is accelerated modestly, from 400 (s11\_01 model) up 1300\,\kms (s11\_2 model). At breakout, the kinetic
   energy of the envelope is a diminishing fraction of the total envelope energy, which is primarily stored internally.
   Asymptotically, the total kinetic energy of the material ejected (a fraction only of the total) is in the range 4$\times 10^{47}$
   up to 10$^{49}$\,erg, with mass-weighted average velocities in the range 70--300\,\kms.
   The long-term light curves are characterized by peak luminosities in the range of 10$^6$--10$^7$\,\lsun\,
   and very long plateau durations of 1--2 years, yielding low-luminosity long-lived transients (Fig.~\ref{fig_s11_lum_all}).
   Hence, these events look somewhat like SNe because they
   attain luminosities/expansion-rates that are comparable to those of low-luminosity/energy Type II-P SNe.
   But they also look like giant eruptions as seen in the most massive
   luminous stars, characterized by super-Eddington luminosities on the order of 10$^7$\,\lsun, with ejected
   shells weighing few solar masses and outflowing velocities of a few hundred \kms.
   The Eddington luminosity for this 10.6\,\msun\, pre-SN star is
   $L_{\rm Edd}= 4 \pi c G M / \kappa \sim 4.07\times 10^5$\,\lsun\,
   (we use a representative mass-absorption coefficient $\kappa=0.34$\,cm$^{2}$\,g$^{-1}$  for this hydrogen-rich
   stellar envelope),
   and hence, the luminosities quoted above in the range 10$^6$--10$^7$\,\lsun\, are all significantly super-Eddington.
   By contrast to radiative-driven {\it outflows}, the super-Eddington nature of the luminosity is only a feature
   in the current context since the {\it ejecta} is already unbound after shock passage (this does not exclude
   secondary effects that could stem from the high luminosity).

   The outcome of such a weak energy ``explosion'' can be two things. If the origin is a very weak energy release following
   core collapse, then the fraction of mass that was expelled will lead to a one-time transient, and the proto-neutron star will accumulate
   the inner part of the envelope, perhaps transitioning to a black hole if it becomes sufficiently massive (in the s11\_01, it would
   form a black hole with a 3.85\,\msun\, baryonic mass). But if the weak energy release is caused by something else, e.g.
   thermonuclear combustion of material at the surface of the degenerate (and hydrostatically-stable) core, the outcome
   is a massive eruption in a massive star. In this case, the luminosity eventually levels off, as the photosphere receding in
   mass through the ejecta layers eventually reaches the inner envelope layer that failed to eject.
   This second category of objects would be called SN impostors.\footnote{In this case, the stellar core would eventually
   collapse, potentially ejecting the residual envelope, which is now of much lower mass and quasi hydrogen-deficient.
   Depending on the time delay between the two events, this could also produce a strong interaction with the previous
   ejected shell, yielding an interacting SN.}

   Unless the event occurs nearby and allows the inspection of the transient site, one does not know whether a
   star survived or not. Hence, such events are ambiguous and may be interpreted either as a one-time
   non-repeatable explosion (a SN) or a potential recidivist like $\eta$ Car.
   These are impostors in the sense that they would not result from the collapse of a degenerate core, but they do share
   that important property with core-collapse SNe that, in the present context,
   they would result from shock-heating of the envelope. Hence, a breakout signal would systematically
   herald the forthcoming ejection, excluding its origin as a steady-state super-Eddington wind accelerated from a hydrostatic base.
   The longer duration of $\sim$1\,d  and the softer  distribution (UV-peaked) of the breakout signal
   should allow the detection of such events in large-scale sky surveys, providing a clear discriminant between
   an explosive or a steady-state-wind origin for such mass ejections.

\begin{table*}
  \begin{minipage}{140mm}
    \caption{Same as Table~\ref{tab_s11}, but now showing the results for the sequence of models with
    differing energy deposition depth in the 11\,\msun\, progenitor star (see \S\ref{var_mcut} for discussion).
\label{tab_s11_wcomp}}
\begin{tabular}{lcccccccc}
\hline
Model Name                                   & s11\_1  & s11\_1\_w3  & s11\_1\_w7 & s11\_1\_w9  &  s11\_3  & s11\_3\_w3 & s11\_3\_w7  & s11\_3\_w9      \\
\hline
$E_{\rm dep}$  (10$^{50}$\,erg)              & 1(-1)   & 1(-1)      &  1(-1)    & 1(-1)      &  4(-1)   &     4(-1) &    4(-1)   & 4(-1)      \\
$M_{\rm cut}$  (\msun)                       & 1.8     & 3.0        &  7.0      &  9.0       &   1.8    &     3.0   &    7.0     &   9.0   \\
$t_{\rm edep}$     (s)                       & 10      & 10         &  10       &  10        &  10      &     10    &    10      &   10  \\
$t_{\rm end}$ (d)                            &  730    & 450        & 450       &  354       &  160     &    200    &    200     &  200    \\
$M_{\rm ejected}$  (\msun)                   & 7.75    & 7.60        & 3.60       &  1.60       &   8.72   &    7.60    &    3.60     &   1.60   \\
$\langle v \rangle_{M}$  (\kms)                            &  121    & 350        & 520       &  804       &  510     &    715    &  1000      & 1600     \\
$(E_{\rm kin})_{\rm end}$  (10$^{50}$\,erg)  &1.3(-2) & 1.0(-1)      & 1.0(-1)     &  1.0(-1)     &  2.8(-1)&   4.1(-1) &  4.1(-1)   & 4.3(-1)     \\
$\langle v_{\rm max} \rangle$  (\kms)                      & 560     & 1200       & 1430      &  1650      &  1800    &   2000    &  2400      & 3200     \\
$t_{\rm SBO}$ (d)                            & 21.4    & 9.7        & 4.3       &  1.8       &  5.8    &    5.0    &  2.2       & 0.9     \\
$L_{\rm peak, SBO}$ (\lsun)                  & 5.2(7)  & 1.0(9)     & 1.8(9)    & 3.7(9)     &  4.9(9)  &    6.9(9) &  1.2(10)   & 2.4(10)     \\
$\langle v_{\rm shock}\rangle$  (\kms)                    & 221     & 447 &     520       &  740       &  812     &    870    &  1000      & 1460     \\
$(E_{\rm kin})_{\rm SBO}$ (erg)              & 6.7(47) & 4.3(48)    & 3.3(48)   &3.0(48)     &  1.3(49) &   1.7(49) &  1.3(49)  & 1.2(49)     \\
$(E_{\rm int})_{\rm SBO}$ (erg)              & 3.8(48) & 6.7(48)    & 7.6(48)   &8.3(48)     &  1.7(49) &   2.6(49) &  3.0(49)  & 3.3(49)     \\
$t_{\rm peak,plateau}$ (d)                   & 243     & 177        & 122       & 80         &  97      &   135     &  96        & 64     \\
$L_{\rm peak, plateau}$ (\lsun)              & 2.9(6) & 1.8(7)    & 3.2(7)    & 4.4(7)     &  2.9(7)  &   5.4(7)  &  9.8(7)    & 1.3(8)     \\
$\Delta t_{\rm plateau}$ (d)                 &   341   & 181 &     126       & 87         &  $>$155  &   142     &  102       &   71    \\
$\int L_{\rm bol}$ dt (10$^{49}$\,erg)       & 2.5(-2) & 7.1(-2)    & 7.7(-2)   &  8.0(-2)     &  1.3(-1) &   1.8(-1) &  2.0(-1)     &  2.2(-1)    \\
\hline
    \end{tabular}
  \end{minipage}
\end{table*}

   \subsubsection{$E_{\rm dep} < E_{\rm binding}$}

   For the smallest energy deposition value in our sequence
   (model s11\_0; $E_{\rm dep}=$5$\times$10$^{48}$\,erg), only a tiny fraction
   of the envelope ($\sim$0.13\,\msun) is ejected, concomitanly with shock breakout,
   while the bulk of the envelope remains attached to the star.
   The energy deposition is used to lift the envelope out of the gravitational potential well, but falling short
   from unbinding it. This results in a ``bloated'' star whose radius
   has increased by about a factor of two in 50 days after breakout, while the surface temperature has hardly changed.
   Consequently, the corresponding star, with its puffed-up envelope,
   brightens from a luminosity of $\sim$10$^5$\,\lsun\, up to $\sim$1.37$\times$10$^5$\,\lsun,
   which remains well below its Eddington luminosity of 4.07$\times$10$^5$\,\lsun.

   Over the two-year time-scale covered by the simulation, the photosphere retains a similar radius
   and temperature. Since the energy left over by the shock is stored at large optical depth, and since
   it was insufficient to communicate any kinetic energy to the envelope, the transport regime is no
   longer of dynamic diffusion (as in the high energy-deposition cases) but instead of static diffusion,
   with characteristic time-scales of $\sim$100\,yr for this massive star envelope (Fig.~\ref{fig_tdiff_mr}).
   Albeit weak, such an energy deposition causes a significant translation of the star in the HR diagram
   and may be linked, in some cases, to the photometric and spectroscopic variability of H-rich massive stars. The central
   element that enables such low-energy perturbations to have any effect is the very low binding energy of massive
   star envelopes, in particular at the low-mass end.

 \subsection{Effect of varying the Energy-Deposition depth}
\label{var_mcut}

\begin{figure*}
\epsfig{file=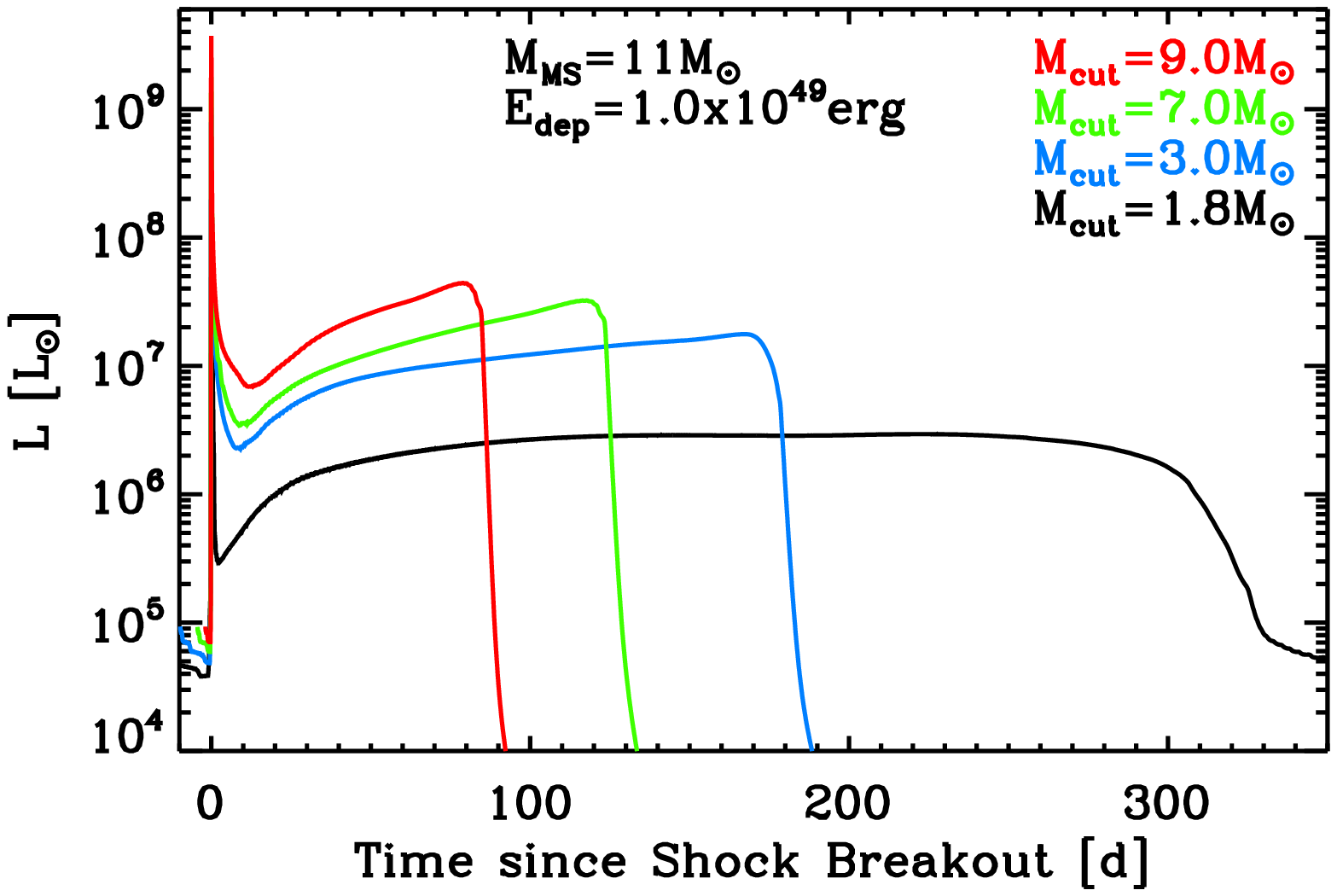,width=8.5cm}
\epsfig{file=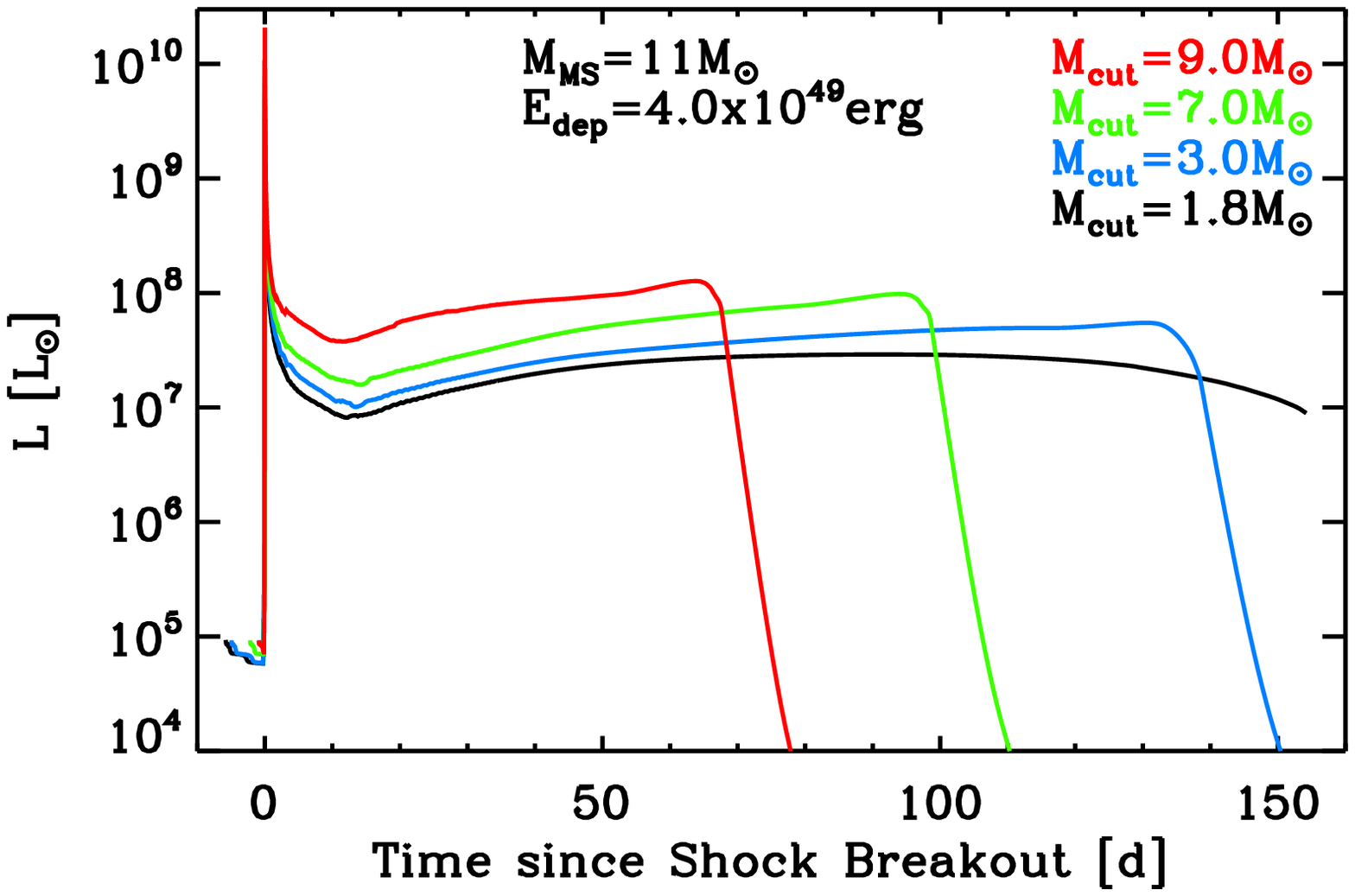,width=8.5cm}
\caption{Same as Fig.~\ref{fig_s11_lum_all} for the 11\,\msun model, but now showing the
emergent bolometric-light evolution resulting for cases in which the energy is deposited
at a Lagrangian mass of 1.8 (black), 3 (blue), 7 (green), and 9\,\msun\, (red).
The left (right) panel corresponds to an energy deposition of 1 (4) $\times$10$^{49}$\,erg.
In all cases, we choose the time origin as the time of shock breakout.
\label{fig_s11_wmin_phot_lum}
}
\end{figure*}

    We have argued in the introduction and in \S\ref{sect_input} that, physically, a deposition of energy from deep regions
    in the progenitor star is more likely since this is where most of the energy is stored or can be made available.
    This can occur through gravitational collapse of the degenerate core from its original radius of a few 1000\,km to $\sim$10\,km
    (the standard initiation of a core-collapse SN explosion), or from violent shell burning above the core (our preferred
    trigger for transient phenomena in massive stars). In contrast, the envelope layers well above the core, characterized by lower
    densities and temperatures, tend to merely react to energy changes occurring at depth. It is, however, interesting to investigate the dependency of our
    results on the adopted site of energy deposition. We have thus performed two sequences for $E_{\rm dep}=10^{49}$\,erg (model s11\_1) and
    $E_{\rm dep}=4\times10^{49}$\,erg (model s11\_3), with a deposition at 1.8 (as before), 3, 7, and
    9 \,\msun\, Lagrangian mass coordinate. These correspond to locations in the helium or in the hydrogen shell.
    The deposition time in all cases is 10\,s. We show the resulting
    light curves for both sequences in Fig.~\ref{fig_s11_wmin_phot_lum}, with the corresponding key parameters
    given in Table~\ref{tab_s11_wcomp}.

    For the s11\_1 model sequence (black curve, left panel of Fig.~\ref{fig_s11_wmin_phot_lum}), the energy deposition is on the order
    of the binding energy of the {\it entire} envelope above the 1.8\,\msun\, mass cut. As we discussed before, this led to
    a SN-impostor-like event, with only a fraction of the total envelope ejected to infinity. The envelope outside of a 3\,\msun\,
    mass cut has, however, only a 10$^{47}$\,erg binding energy so, as we move outward the site of energy deposition (from 3,
    to 7, and 9\,\msun\, Lagrangian mass coordinate), we switch to a successful explosion with sizable kinetic energy and rapid light-curve
    evolution.
    As given in Table~\ref{tab_s11_wcomp},  the models with a mass cut at and beyond 3\,\msun\, lead to full ejection of the layers exterior
    to the energy-deposition site, with representative ejecta expansion rates of 400--800\,\kms and kinetic energy of 10$^{49}$\,erg.
    Interestingly, the phase of high-brightness lasts from 90 up to 180 days despite the low amount of ejected material, i.e. between
    1.6 and 7.6\,\msun. The long duration of these events, even for low ejected masses, is caused by two factors. First, the expansion is relatively slow so
    that radiative diffusion is modestly facilitated by the ejecta density reduction with time (we are closer to a static regime of radiative diffusion).
    Second, the energy losses through expansion are minimal since we start from a very extended (i.e. loosely-bound) stellar envelope.
    Note here that in the more compact (i.e. tightly-bound) progenitor configurations for Type I SNe (Wolf-Rayet stars or white dwarfs),
    the cooling through expansion is tremendous and the events owe their large brightness at a few weeks after explosion
    entirely to the heating caused by unstable isotopes with week-long half-lifes. {\it A corollary is that weak-energy ejections with little $^{56}$Ni
    can only reach a large luminosity if they arise from a loosely-bound initial configuration, e.g. a big star.}

    For the s11\_3 model sequence (light curve shown in the right panel of Fig.~\ref{fig_s11_wmin_phot_lum}), the energy deposition is greater than
    the binding energy of the envelope above the 1.8\,\msun\, mass cut so that no matter what the position of the mass cut
    is outside of 1.8\,\msun, we obtain full ejection of the shocked layers, with representative velocities in the range 500-1600\,\kms.

    To conclude, we find that moving the energy-deposition mass cut $M_{\rm cut}$ outwards for a given energy-deposition magnitude
    $E_{\rm dep}$ produces a similar effect to increasing $E_{\rm dep}$  at a given $M_{\rm cut}$. Quantitatively, this is modulated by
    the binding energy of the envelope exterior to $M_{\rm cut}$. We observe in particular that with such RSG progenitors
    it is possible to obtain light curves with
    a very extended plateau even for very small ejecta mass. Note that this remains an experiment though, since we believe the energy
    deposition in a massive star must occur in the vicinity of the core, hence
    interior to the Helium shell of pre-SN massive stars.

\begin{figure}
\epsfig{file=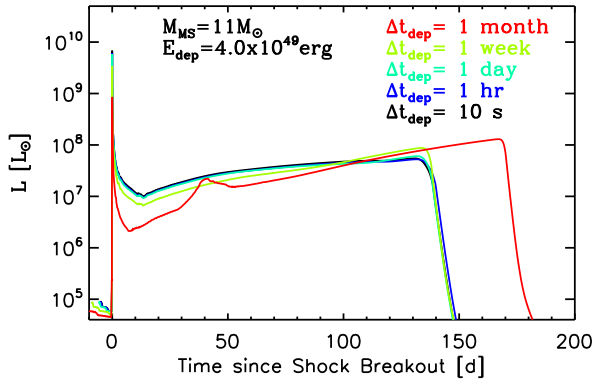,width=8.5cm}
\caption{Same as Fig.~\ref{fig_s11_lum_all}, but now showing the
bolometric light evolution resulting from models in which the energy is deposited at a fixed rate
over 10\,s (black), 1\,hr (blue), 1\,d (light green), 1\,week (green), and 1 month (red).
The site of energy deposition is at Lagrangian-mass coordinate 3\,\msun\, and the energy deposition
is 4$\times$10$^{49}$\,erg in all cases. The time origin is the time of shock breakout.
\label{fig_var_dt}
}
\end{figure}

   \subsection{Effect of varying the Energy-Deposition duration}
\label{var_dt}

   A fundamental element of our study is that energy deposition must occur on a short time-scale
   to ensure the formation of a shock, rather than on a long time-scale which would allow radiative
   diffusion and convection to carry outward the extra energy. In this section, we present results
   with V1D for a model with $E_{\rm dep}=4\times 10^{49}$\,erg and $M_{\rm cut}=3$\,\msun,
   but an energy-deposition time of 10 seconds, one hour, one day, one week, and one month.
   Since the resulting ejecta and bolometric properties are similar to those of the model s11\_3\_w3
   (see Table~\ref{tab_s11_wcomp}), we do not tabulate the results for this sequence of models.
   However, we show the synthetic light curves in Fig.~\ref{fig_var_dt}.
   In all cases, a phase of shock breakout takes place, followed by a bright plateau that persists over a similar
   duration. No qualitative difference is visible, and quantitative differences are small.

   Hence, shock formation persists even for very long deposition times, despite the large range covered.
   In other words, a deposition of energy that takes place over a time-scale much longer than the expected shock revival
   time in core-collapse SN explosion (i.e. one second) still leads to shock formation. We are aware that
   the neglect of feedback effects makes this exploration somewhat artificial. In the context of a nuclear
   flash, expansion could, for example, endanger the continuation of nuclear burning.

   For energy-deposition time-scales much longer than a month, a diffusion wave, rather than a shock, would form,
   together perhaps with convection. Such a regime, not of interest in the present study,
   would have to be followed in 2D or 3D (see, for example, \citealt{mocak_etal_08,mocak_etal_09})
   to properly model the interplay between these two means of energy transport and the resulting effects
   on the envelope structure.

\begin{figure}
\epsfig{file=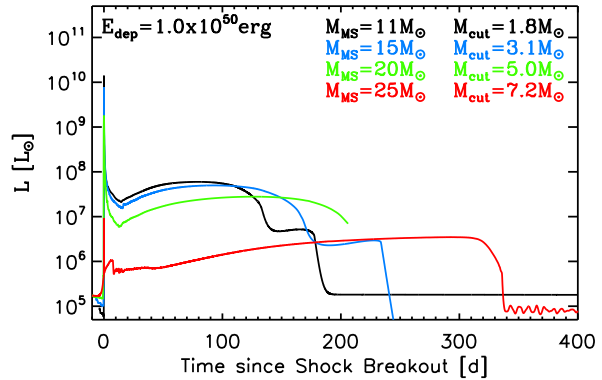,width=8.5cm}
\caption{Same as Fig.~\ref{fig_s11_lum_all}, but now showing the
bolometric light evolution for different progenitor-mass models:
11 (black), 15 (blue), 20 (green), and 25\,\msun\, (red; these masses correspond
to main-sequence masses).
The corresponding total progenitor masses at the time of collapse for this set is
10.6, 12.6, 14.7, and 12.5\,\msun, respectively. In all models,
we deposit the same energy of 10$^{50}$\,erg, and at the same {\it physical} location
in the envelope, corresponding to the base of the helium shell. In the same order,
this corresponds to the Lagrangian-mass coordinate of 1.8, 3.1, 5.0, and 7.2\,\msun.
The time origin is the time of shock breakout, which corresponds to a range of delays
since energy-deposition in the four models.
\label{fig_var_mprog}
}
\end{figure}

  \subsection{Dependency on Progenitor Mass}
\label{var_mprog}

  The choice of the 11\,\msun\, model for the investigations presented so far was motivated
  by the lower density of the progenitor envelope. This produces larger Courant-time increments
  for the hydrodynamics and easier computing. However, as evidenced by the variation in
  envelope binding energy for 10-40\,\msun\, massive stars (recall that these model masses correspond
  to those on the main sequence; the corresponding masses at the onset of collapse can be seen from
  Fig.~\ref{fig_rho_mr}) and by the dependencies presented
  in the previous sections, a given energy deposition will obviously produce a very different outcome
  for different progenitor masses. In this section, we thus explore the
  radiative and dynamical properties of ejecta produced by the deposition of 10$^{50}$\,erg at the base of the helium
  shell in 11, 15, 20, and 25\,\msun\, massive-star progenitors. In the same order, this corresponds
  to mass cuts at 1.8, 3.1, 5.0, and 7.2\,\msun\, in the envelope, to envelope binding energies (we quote
  absolute values) of 9.9$\times 10^{48}$\,erg, 4.6$\times 10^{49}$\,erg, 6.6$\times 10^{49}$\,erg,
  and 1.4$\times 10^{50}$\,erg, and to total progenitor masses at the time of collapse of
  10.6, 12.6, 14.7, and 12.5\,\msun. We show the resulting light curves in Fig.~\ref{fig_var_mprog}.

 As before with the exploration of outcomes for varying energy-deposition magnitudes (\S\ref{var_edep})
  or sites (\S\ref{var_mcut}), adopting different progenitor stars and depositing the energy
  in regions that are more or less bound leads to a similar modulation in light-curve properties.
  With the adopted value $E_{\rm dep}=$10$^{50}$\,erg, we obtain light curves reminiscent of
  low-luminosity Type II-P SNe if it is larger than the binding energy of the overlying envelope
  (11 and 15\,\msun\, models), and long-lived and fainter events reminiscent of SN impostors
  or low-energy transients otherwise  (20 and 25\,\msun\, models).

  Note that if, using the same progenitors (11, 15, 20, and 25\,\msun\, main-sequence mass models),
  we deposit a given energy at a radius where the Lagrangian mass is 2\,\msun\, lower than the total
  progenitor mass (same shell mass underneath the progenitor surface), we obtain essentially
  the same light curves and ejecta properties for all models. This degeneracy results from the generic
  $\sim$10$^{47}$\,erg binding energy of the corresponding (outer) H-rich shell in these different stars.

  To summarize, and perhaps paradoxically, the consideration of these additional progenitors
  (with different envelope structure/binding energy) does not add any new perspective to our results so far since
  the diversity of outcomes was fully revealed earlier on with the 11\,\msun\, model through the variation of
  the values of $E_{\rm dep}$ and $M_{\rm cut}$.

  Our sample of progenitor models does not include more massive
  stars. Such models were not added to retain a homogeneous sample, focus on conceptual aspects (i.e. shock heating
  of a representative gravitationally-bound stellar envelope) and leave
  aside the specificities of the numerous different categories of stars undergoing eruptions.
  One such category is that of blue-supergiant stars in the phase of hydrogen/helium core/shell burning.
  These objects are found at moderate masses, covered by our sample, but also extend to much larger
  masses. For example, the stars associated with events as diverse as SN1987A and the erupting $\eta$ Car 
  are BSGs, with associated masses in the range 15-20\,\msun\ and $\sim$100\,\msun.
  However, compared to our sample of RSG stars, these objects are all more tightly bound,
  as reflected by their smaller surface radii of at most a hundred rather than a thousand times that of the sun.
  Hence, the various regimes highlighted above would apply with all energies scaled upwards.
  The duration of the transient would depend on the amount of material ejected, reflecting the
  dependencies discussed in \S\ref{var_mcut}.
  Because of their larger binding energy, very low-energy transient resulting from the mechanism presented
  here would be less likely to occur in such objects.  For the most massive objects, the proximity to
  the Eddington limit will considerably reduce the binding energy of the envelope and will make it more
  prone to eruptions if perturbed by a deep-seated energy source. To provide more quantitative results and comparison
  with observations, we will present in a forthcoming study the gas and radiation properties we predict for
  very massive post main-sequence stars subject to shock heating.

\section{Sample of Non-LTE Synthetic Spectra at 5 Days after Shock Breakout}
\label{sect_rad}

    In this section, we provide synthetic spectra that illustrate the spectral-energy distribution
of the various models discussed above. We choose the representative time of five days after
shock breakout since it is a typical discovery time for SNe, and perhaps as well for
optical transients etc. This also offers a unique reference time for comparison in all models.
We focus on the models presented in \S\ref{var_edep} in which only the energy deposition
magnitude is varied, using the 11\,\msun\, model and adopting a deposition site at the 1.8\,\msun\,
Lagrangian-mass coordinate.

\begin{figure*}
\epsfig{file=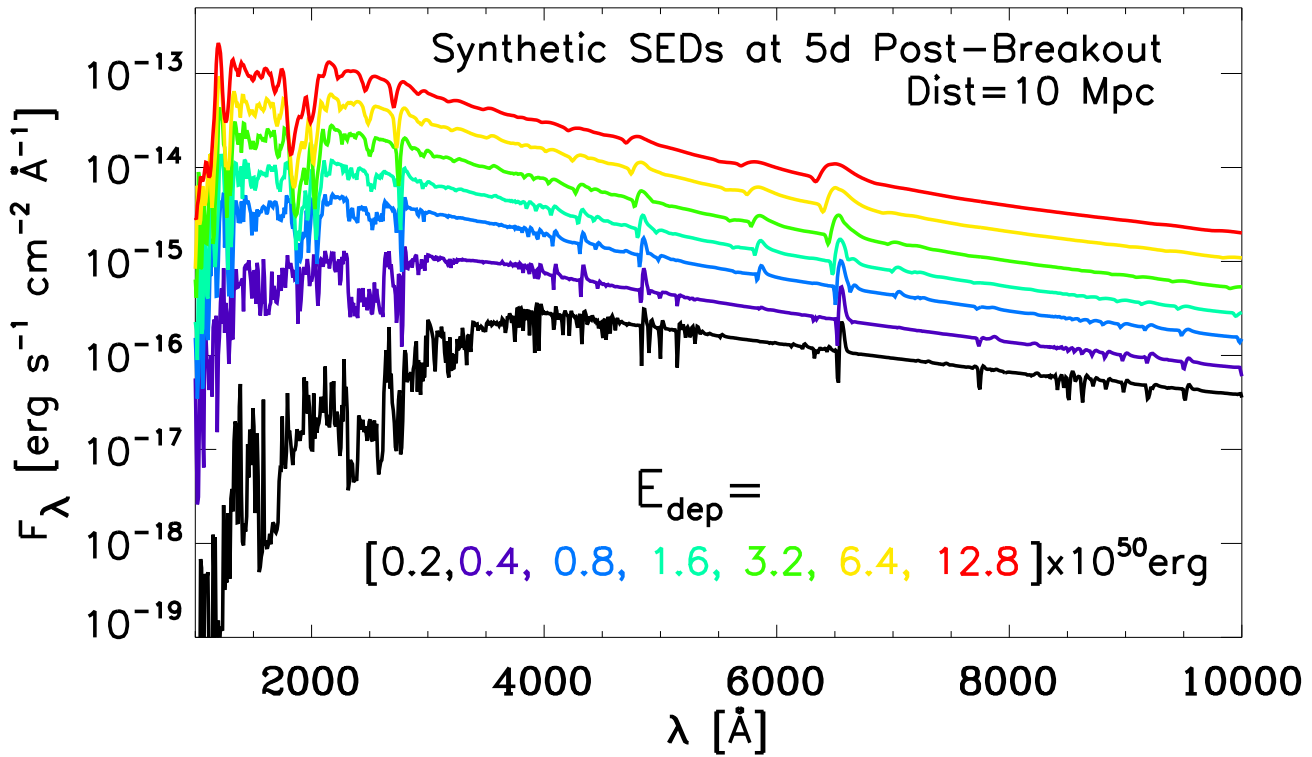,width=18cm}
\caption{Non-LTE synthetic spectra for the models presented in \S\ref{var_edep}, based on the
ejecta properties computed with V1D and extracted at a representative time of 5 days after shock breakout
(we scale the synthetic fluxes adopting a distance to the object of 10\,Mpc).
Notice how, despite the 2-order-of-magnitude range in energy deposition in this model sequence,
the resulting spectral-energy distributions are
qualitatively similar, differing essentially in the absolute flux level (the larger the energy deposition,
the more luminous the event) and in line-profile width (the larger the energy deposition, the larger the
ejecta kinetic energy). This similarity is also seen in observations of hydrogen-rich transients
and results from the comparable envelope composition of the progenitor star and the comparable
ionization state at the ejecta at the photosphere. \label{fig_v1d_sed}
}
\end{figure*}

   The computation we perform follows the approach described in various contexts in
\citet{DH06_SN1999em,dessart_etal_08,dessart_etal_09}. Using the results from our V1D
computations, we extract the gas properties at a total optical depth of a few tens and at a time
of about five days after shock breakout. We then use these characteristics of radius, density,
temperature, and velocity (see Fig.~\ref{fig_s11_phot})
to setup the initial conditions for the non-LTE radiative-transfer calculation.
At such early times, the photosphere resides in the outer hydrogen-rich part of the progenitor
envelope, characterized by a typical RSG surface composition (see, e.g., \citealt{dessart_etal_08}).
We adopt a density exponent of ten and a homologous expansion. Recall that the line and continuum
formation regions are rather confined and hence such characteristics, in practice, really matter {\it locally}, i.e.
the density/velocity distributions may vary differently in a global sense (for details, see \citealt{dessart_etal_09}).
We show in Fig.~\ref{fig_v1d_sed} the synthetic spectra (i.e. the emergent flux in the observer's frame), assuming
a distance of 10\,Mpc and neglecting any reddening.

At that reference time, and from the high to the low energy-deposition case, we obtain a trend towards smaller
ejecta velocity (narrower line profiles), smaller temperature (redder peak flux distribution), combined with a smaller radius
(smaller flux). At higher energy deposition, the synthetic spectra are reminiscent of Type II-P SNe (from
standard to low-luminosity ones), while for lower energy deposition, the spectra are qualitatively similar
to those of cool mass-losing stars like the present day $\eta$ Car or SN
impostors, e.g. 1961 \citep{zwicky_61}, 1997bs \citep{vandyk_etal_00}, 2000ch
\citep{wagner_etal_04}, or  2002kg \citep{maund_etal_06}. We thus recover spectroscopically
the various regimes identified in bolometric light curves and ejecta properties, primarily through
modulations in luminosity and line width.
 An interesting aspect of this set of simulations is that for H-rich massive stars (and as long as the temperatures
   are not too low to foster the formation of molecules and dust), we obtain a very uniform set of spectral features
   (with the dominance of hydrogen and metals, whose abundance is that of the environment in which the star
   was born), merely varying in widths, with slight changes in the flux distribution mostly associated with variations
   in ionization. For example, the LBV $\eta$ Car has the same surface composition as the progenitor of
   SN 1999em \citep {hillier_etal_01,DH06_SN1999em}; it also has similar line-profile morphology and
   flux distribution as that of the interacting SN 1994W \citep{dessart_etal_09}.
   This spectroscopic degeneracy/uniformity adds to the difficulty of inferring the dynamical origin of the event.

\section{Discussion}
\label{sect_discussion}

    In this section, we discuss the ramifications of our results for explosions and eruptions in massive stars.
Let us stress again that we do not make the claim that all massive-star mass ejections, or all transients in general, stem from a strong, sudden,
and deeply-rooted energy release in the progenitor-star envelope. But here, given this working hypothesis, we explore
its potential for explaining the properties and the diversity of observed transient phenomena associated
with massive stars.

\subsection{Binding energy considerations}

 The binding energy is the most fundamental quantity controlling the outcome of a given energy deposition.
Although this conclusion seems trivial, its implications are far-reaching.

 Since the binding energy of massive stars increases with progenitor mass, SN explosions arising
 from more massive progenitors will, on average, be more energetic. Indeed, the energy deposition at the base
 of the envelope to be ejected has to be a few units of the binding energy, one unit to unbind the envelope,
 and the remaining units to give it its kinetic energy (fast expansion) and internal energy (bright display).
 In the sample of massive stars studied here, the envelope binding energy (right panel of Fig.~\ref{fig_eb_mr})
 sets a higher energy threshold for increasing progenitor main-sequence mass. Consequently, stars originally
 more massive than $\sim$15\,\msun\, require an energy deposition of at least 10$^{51}$\,erg to unbind the envelope
 while 100 times less is sufficient in a 10\,\msun\, progenitor.

 A corollary is that quite generally, low-energy explosions in gravitationally-bound objects
 are more likely to be associated with loosley-bound objects. This appears counter intuitive
 since one may argue that the ejecta kinetic energy at infinity should be anything from zero
 ($E_{\rm dep} = E_{\rm binding}$) up to arbitrary large values (set by $E_{\rm dep} - E_{\rm binding}$).
 This is true in theory, but in practice, the range of energies for events classified as a SN or as a low-energy
 transient differs, and vary greatly with the progenitor binding energy.
 At present, we have no reason to believe that $E_{\rm dep}$ and  $E_{\rm binding}$ are exactly
 tuned in any way. This holds very obviously in the context of the core-collapse SN explosion mechanism where
 the conditions for a successful explosion are set by what happens in the vicinity of the core and thus entirely
 disconnected from the {\it global} threshold set by the envelope binding energy. So, we can assume that
 $E_{\rm dep}$ is {\it uniformly} distributed (the number of events having an energy between
 $E_{\rm dep}$ and $E_{\rm dep} + \Delta E$ scales linearly with the energy bin $\Delta E$ but is
 independent of $E_{\rm dep}$), with values above and below the envelope-binding energy.
 Now, mass ejection will only occur if $E_{\rm dep} >  E_{\rm binding}$. Consider an object with a binding
 energy of 10$^{51}$\,erg, and focus on the {\it uniform} distribution of events with an energy deposition between
 10$^{51}$\,erg and 2$\times$10$^{51}$\,erg (we ignore those events that lead to no eruption and limit the maximum
 energy to that for a typical SN ejecta, e.g. for SN1987A). We can say that all ejecta with a kinetic energy in the range
 10$^{50}$\,erg up to 10$^{51}$\,erg represent powerful explosions and will tend to be viewed/classified as
 a SN: This represents 90\% of the events in our distribution. Ejecta with an (asymptotic) kinetic energy less than 10$^{50}$\,erg
 will be produced with a 10\% probability, those with less than 10$^{49}$\,erg with a 1\% probability, and those
 with less than 10$^{48}$\,erg with a 0.1\% probability. This does not mean that a low-energy transient cannot take place
 in a highly bound object, but it strongly suggests that unless there is a mechanism that biases the energy deposition
 to match the binding energy (all SN explosions strongly support the contrary!), low-energy ejecta are {\it statistically
 more likely} to occur in loosely-bound objects.  Another way to consider this is that to produce a low-energy
 transient in a highly-bound object, the excess energy to deposit is
 so small compared to the binding energy of the envelope that the required tuning to
 make this low-energy ejection can hardly be met.
 Hence, low-energy transients associated with stars should stem from ejection of loosely-bound regions, for example
 the envelope of low-mass massive stars or, in another context, low-mass shells at the surface of a white dwarf.
 Obviously, this is relevant only if any such energy deposition can take place but observations of transients
 demonstrate that it can. How it does it  is a question left for future work.

 As more and more energy is needed to eject the envelopes of more massive progenitors, the extreme case of tightly
 bound Wolf-Rayet stars can only be associated with very energetic explosions. The fact that (nearly) all hypernovae are associated
 with Type Ib/c SNe is in this respect quite compelling. It is not so much that they somehow
 can achieve higher energy deposition; it is that they {\it must} (the fact that they do achieve a higher energy-deposition
 to explode is an interesting problem that concerns the explosion mechanism, neutrino physics, magneto-rotational effects etc).
 In other words, of all possible scenarios, {\it only} those
 that lead to a very large energy deposition (well in excess of the binding energy) can yield a successful, and visible, explosion.
 This may act as a natural tuning ingredient to favor fast-rotating cores in tightly-bound progenitors like Wolf-Rayet stars
 \citep{burrows_etal_07b}. The fraction of Wolf-Rayet stars that fail to explode following core collapse is unknown at present.

  This binding-energy argument may also be a very natural explanation for the lack of detection of Type II
  SN explosions for objects with main sequence mass greater than $\sim$20\,\msun\, \citep{smartt_etal_09}.
  The standard neutrino mechanism may, after all, fail to deliver an energy that is comparable
  to the binding energy of the more massive progenitor stars. And indeed,  this may not be so surprising given the
  ten order of magnitude density contrast in the region 2-5\,\msun\, between the 11 and the 25\,\msun\, models of
  WHW02 (Fig.~\ref{fig_rho_mr}). This would suggest that a lot of core-collapse events associated with the higher mass massive stars
  do not yield explosions, but instead form a black hole with no transient electromagnetic display (see \citealt{smartt_etal_09}).
   The proposition of  \citet{kochanek_etal_08} to monitor a large number of red supergiants for failed explosion
    would nourish this option.

   It is the low binding energy (hence extended envelope structure) of massive stars that causes the bright
displays of their ejected envelopes, even for low energy deposition,
and even in the absence of any heating from unstable isotopes. In general,
the change from a modest-size progenitor star to an extended  ejecta
causes a dramatic cooling, primarily due to expansion. And indeed, in Type Ia/b/c SNe,
this cooling is dramatic and the large SN luminosity results only because of the presence of unstable isotopes,
whose half-life of days to weeks makes the SN bright on a comparable time-scale. In contrast,
only modest radial expansion (typically by 2 orders of magnitude compared to 5--7  in Type I SNe) occurs after the explosion
of the already extended envelopes of H-rich massive stars. In such objects,
large luminosities of 10$^7$--10$^8$\,\lsun\, can be achieved without the contribution from unstable isotopes,
and even for modest ejecta kinetic energies.

  Hence, in general, low-energy (stellar) explosions/eruptions should occur from regions that are weakly gravitationally-bound.
  Given the very low envelope binding energy of low-mass massive stars and hence the very low energy threshold
  for producing explosion/eruption,
  we argue that these should be prime candidates for transient events in the Universe, perhaps even in connection to interacting SNe.
  The stellar initial-mass function is also biased in favor of such low-mass massive stars, which are therefore not rare (about one for every
  200 solar-type stars).
  This argumentation likely applies to SN 2008S \citep{prieto_etal_08, smith_etal_09b,boticella_etal_09}, M85-OT2006-1
  \citep{kulkarni_etal_2007_m85,pastorello_etal_07_m85}, or the enigmatic SN2008ha \citep{foley_etal_09,valenti_etal_09}.
  As argued above, the association of a low-energy low-luminosity ejecta with a Wolf-Rayet progenitor,
  believed at present to always be a highly-bound object, seems unlikely (although not excluded in theory) in the context 
  where the energy deposition occurs deep
  within the progenitor star. In that context, it would require the presence of a loosely-bound shell at the surface of the progenitor
  star and a pretty exquisite tuning to allow the energy deposition occurring at depth to be no more than a few 1.001 times
  the envelope binding energy. Some of these postulates may not apply (e.g. could the energy deposition occur just below the
  surface of the WR star?), but it is unclear at present which is faulty.
  Specific radiation-hydrodynamical simulations based on suitable progenitors are needed to address this problem.

\subsection{Explosion/eruption associated with transient phenomena}

   A growing sample of transient objects possess intermediate luminosities between SNe and novae.
   Their origin is debated. Owing to their low energy, the community has argued for either a weak-energy core-collapse
   SN explosion (e.g. \citealt{boticella_etal_09,pastorello_etal_09}), or for a super-Eddington wind in a SN impostor
   (e.g. \citealt{maund_etal_06,smith_etal_09}). The ambiguity stems from the common energy-ground in which events
   associated with these two scenarios fall, the $\lesssim$10$^{50}$\,erg kinetic energy of the homunculus
   associated with $\eta$ Car rivaling that of low-luminosity Type II-P SN ejecta. Reversing the argument, how many of these
   low-luminosity Type II-P SNe are in fact impostors, boasting ejecta kinetic energy that are well below that of
   LBVs? The ``SN'' status in this energy regime seems to be often based on faith rather than on unambiguous observational or
   theoretical evidence.

   In the shock-heated solutions we presented in \S\ref{var_edep},
   the ejecta are systematically turned into a radiation-dominated plasma. At shock breakout, they are essentially
   unbound (their total energy $E_{\rm grav} + E_{\rm kin} + E_{\rm int} $ is positive) and possess a large amount of internal energy.
   In those models with low-expansion velocity and super-Eddington luminosities of 10$^6$--10$^7$\,\lsun, the photosphere
   has properties that are comparable to those of erupting LBVs, yet their super-Eddington luminosity plays
   no role in driving the material, which was already unbound after shock passage.
   From an inspection of the long-term light curve, it seems difficult to distinguish between the two
   since the time-scale over which the radiative display evolves is very long and related to radiative diffusion:
   Apart from the breakout signature, it bears no imprint of the explosive origin, i.e. the sudden and deeply-rooted
   release of energy.

   The recent observation of fast-moving material ahead of $\eta$ Car's homunculus
   points very strongly toward a shock-heated envelope as the origin of the giant eruption \citep{smith_08_blast}.
   The invocation of a super-Eddington wind in the low-mass progenitor star associated with SN2008S may be supporting
   the same hypothesis since low-mass massive stars are well below their Eddington limit, while
   a shock could easily take them beyond that limit.
   An interesting and generic component of all the shock-heating scenarios that we explored in this paper
   is the associated shock-breakout signal, whose detection could therefore disentangle situations associated with a quasi-steady wind
   from a hydrostatically-bound super-Eddington atmosphere on the one hand from
   those associated with a slowly-moving shock-heated ejecta on the other hand.

\subsection{Variability in Massive stars}

   While the binding energy of massive star envelopes increases with progenitor mass, the mean binding
   energy is low for all objects that have retained a hydrogen envelope. In particular, and in the models
   of WHW02, that outer H-rich shell has a very uniform and very low binding energy of 10$^{47}$\,erg.
   Owing to mass loss, the mass of that shell decreases with time for higher-mass massive stars,
   but it always remains loosely bound. In our simulations, we found that even for a very modest energy
   deposition, well below the binding-energy value, important structural changes of the envelope
   occurred.

   We suspect that hydrodynamic-fluid instabilities taking place deep in the stellar interior
   \citep{bazan_arnett_94,bazan_arnett_98,asida_arnett_00,arnett_etal_05,meakin_arnett_06}
     may provide the energy seed for at least a fraction of the variability observed in massive stars.
   This has not been fully realized so far in part because studies of stellar variability tend to isolate the surface
   layers and ignore the interior, while studies of stellar interiors reduce all the complicated
   physics occurring near the surface to imposed and simplified conditions at the grid outer boundary
   (see, however, \citealt{cantiello_etal_09}).
   We expect that, owing to their huge surface radii and therefore low binding energy,
   supergiant and hypergiant stars should be very sensitive
   to such perturbations from the stellar interior, and indeed observations of such stars reveal
   a rich and violent mass-loss history \citep{smith_etal_09}. Similarly, objects that through
   the course of their {\it quasi-steady} evolution happen to lie close to the Eddington limit and/or reach
   critical rotation are naturally more exposed to potential {\it time-dependent} deeply-rooted energy leaks.

    This does not exclude the role of other mechanisms for mass ejections, for example,
   pulsations, rotation, radiation-driven mass loss etc., although it is at present not clear how these
   processes can produce the extreme properties required in the context of high-luminosity
   interacting SNe or giant eruptions of LBVs.

\section{Conclusions}
\label{sect_conclusion}

    In this paper, we have presented one-dimensional one-group (gray) two-temperature radiation-hydrodynamics
simulations of pre-SN massive-star envelopes subject to a sudden release of energy above their degenerate core.
Although at the high energy end, the likely phenomenon is core collapse, we more generally have in mind
the thermonuclear incineration of intermediate-mass elements present in shells above the core. The motivation
for this study is: 1) The existence of interacting SNe (which must eject a shell at least once before core-collapse, but
more likely eject multiple shells, by some means not elucidated so far); 2) the identification of fast-moving material
exterior to $\eta$ Car's homunculus (which cannot stem from radiation-driving in a wind;  \citealt{smith_08_blast});
3) the broad range of
energies inferred for Type II-P SNe, overlapping at the low-energy end with high-energy transients and massive-star
eruptions. The bulk of our results stem from work on the 11\,\msun\, model of WHW02, although tests with higher-mass
progenitors yield the same qualitative conclusions and regimes, merely shifted quantitatively.
 This work is an exploration of the outcome of a strong, sudden, and deeply-rooted energy deposition taking place
at the base of a massive-star envelope. We are not proposing this scenario holds for all massive-star
mass ejections, but we investigate what results when this circumstance obtains. There is no doubt it does,
but how frequently and how robustly is left for future study.

\begin{figure}
\epsfig{file=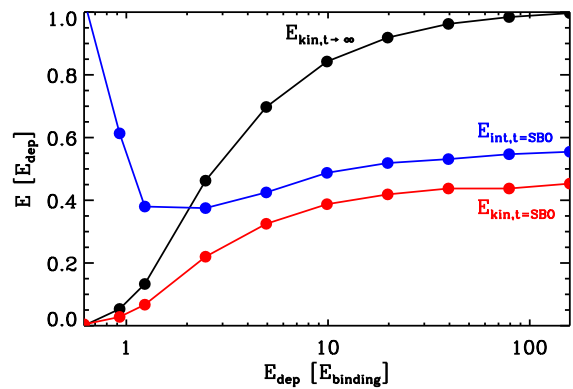,width=8.5cm}
\caption{Correlation between various energies (normalized to the
corresponding adopted energy deposition) and the energy deposition (normalized to the
corresponding envelope binding energy). In black, we show the normalized asymptotic energy of the ejecta/outflow,
and in blue (red) the normalized envelope/ejecta internal (kinetic) energy at the time of shock breakout.
Only the models presented in Table~1 are shown here.
The black curve illustrates the three regimes we have presented here, with SN explosions,
SN impostors, and variable/perturbed stars as the energy deposition varies from being much larger,
comparable, and smaller than the envelope binding energy.
Notice how the internal energy always dominates over the kinetic energy at the time of shock breakout
in the simulations performed here.
\label{fig_summary_ekin}
}
\end{figure}

   Although this result is not new, we find that the fundamental quantity controling the outcome
   of a strong, sudden, and deeply-rooted energy deposition is the binding energy of the stellar envelope,
   which increases considerably with progenitor mass
   in the range 10--40\,\msun. What is new, however, is our study of the long-term evolution of the gas and radiative properties
   that result from configurations where the energy deposited is greater, on the order of, or
   less than the envelope binding energy. We identify three regimes, with a continuous progression from
   1) SN explosions at high energy ($E_{\rm dep} > E_{\rm binding}$), with a complete envelope ejection
   associated with a 100-day long high-plateau luminosity;
   2) SN impostors at the intermediate energy range ($E_{\rm dep} \sim E_{\rm binding}$), with a
   partial envelope ejection, and a more modest but longer-lived plateau luminosity;
   and 3) bloated/variable stars at the low-energy end ($E_{\rm dep} < E_{\rm binding}$), with little or no mass
   ejection but with a residual puffed-up envelope of weakly-modified properties.
   What conditions the results is not the magnitude of the energy deposition itself but how it compares with
   the binding energy of the envelope exterior to the site of deposition. Hence, to achieve the same result requires
   more energy in a more massive progenitor star. These properties are summarized in Fig.~\ref{fig_summary_ekin}.

   In all simulations presented, a shock forms systematically with a strength that depends on the energy deposited.
   It crosses the envelope in  $\sim$1 to $\sim$50 days, hence communicating its energy to the entire progenitor envelope
   {\it quasi-instantaneously}, i.e. as opposed to, e.g., a diffusion time-scale for energy transport of 10$^4$ years or more
   at the corresponding depth. This shock eventually emerges at the progenitor
   surface with a breakout signal that varies from a duration of an hour up to a few days
   (modulated here by the shock speed rather than by the atmospheric-scale height), with a flux peaking
   at longer wavelengths for weaker shock strengths. This breakout signal is the (and may be the only)
   {\it unambiguous} evidence that the subsequent ``ejection" was triggered by
   shock-heating, and thus has an explosive origin.

\begin{figure}
\epsfig{file=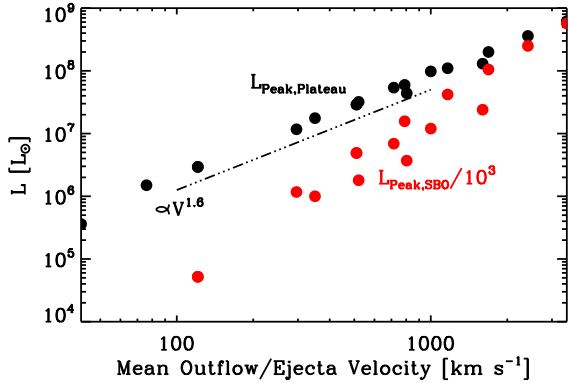,width=8.5cm}
\caption{Correlation between the peak luminosity during the plateau phase (black dots)
and the mass-weighted average ejecta velocity. We also show the correlation for
the peak luminosity at shock breakout (red dots; scaled down by a factor of 1000 for convenience).
Note that we include models from Tables~1 and 2. We overplot the line $L \propto v^{1.6}$, which
follows closely the distribution of points for $L_{\rm Peak, Plateau}$ versus $\langle v \rangle_M$.
Our radiation-hydrodynamics simulations support the correlation identified by \citet{hamuy_pinto_02}
and subsequently improved upon by \citet{nugent_etal_06,poznanski_etal_09}. Our slope is in
close agreement with that proposed in this last reference.
Impressively, the relation holds over the entire domain explored. Note that there is no consideration of
radioactive decay from unstable isotopes or departures from spherical symmetry, and only data points
associated with the 11\,\msun-progenitor star are used. Relaxing these choices would likely introduce some scatter.
\label{fig_summary_lpeak}
}
\end{figure}

\begin{figure}
\epsfig{file=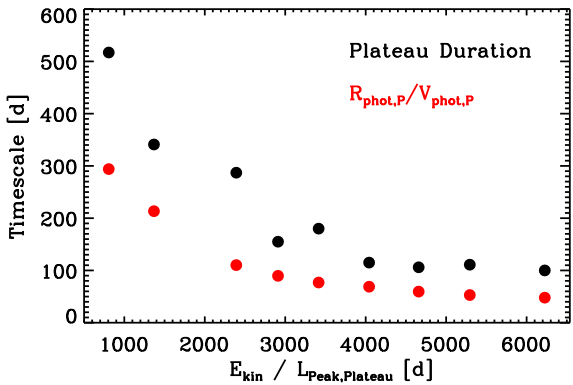,width=8.5cm}
\caption{Correlation between the plateau duration (black) and the time-like quantity $R_{\rm phot,P}/V_{\rm phot,P}$
(ratio of the radius and the velocity at the photosphere at the time of peak-plateau brightness) versus
the time-like quantity equal to the asymptotic ejecta kinetic energy divided by the peak-plateau luminosity brightness.
\label{fig_summary_dt}
}
\end{figure}

   At shock breakout, the luminosity reaches a peak, then fades, before stabilizing/rising again forming an
   extended plateau. This plateau phase corresponds to the recombination epoch of the ejected mass,
   the internal energy decreasing primarily through expansion and little through radiation. It is the large stored
   internal energy (in the form of radiation) that keeps the ejecta optically thick and luminous (radioactive
   decay is neglected in our work). We find a continuum of light curves, faster-expanding ejecta
   being more luminous both at breakout (stronger shock) and during the plateau phase (Fig.~\ref{fig_summary_lpeak}).
   The models presented in \S\ref{var_edep} corroborate the correlation between
   plateau luminosity and  mid-plateau photospheric velocity identified by \citet{hamuy_pinto_02}, and
   refined by \citet{nugent_etal_06,poznanski_etal_09}. We also find that the plateau duration is anti-correlated
   with energy deposition (Fig.~\ref{fig_summary_dt}). At larger energy, faster expansion leads to faster cooling and recombination so that the
   ejecta photosphere recedes faster in mass/radius after reaching its peak earlier. For small energy variations
   in this regime, interplay between kinetic and internal energy (which are comparable at breakout) yield
   a plateau duration that is $\sim$100\,d, which is on the order of Type II-P plateau lengths.
   For lower energy deposition, we switch slowly from a regime of dynamic diffusion to that of quasi-static diffusion.
   The more slowly-expanding ejecta, characterized by a slowly-decreasing
   optical depth with time, gives the bolometric luminosity a modest peak value and a slow evolution,
   with plateau durations of up to 1-2 years.  The plateau phase is thus more extended
   and fainter for lower energy deposition, echoing the light-curve trend going from SNe to SN impostors (Fig.~\ref{fig_obs_lc}).
   Note that, in our simulations, these time-scales are always 6-7 orders of magnitude larger than the time-scale
   for the energy deposition, which was chosen
   to be ten seconds in most cases. In other words, the time-scale over which the light curve evolves (basically that
   of radiative diffusion in an expanding medium) has no connection
   to the time-scale over which the energy was deposited in the first place.

   From our exploration and with our set of models, we find that explosions of varying strength can yield the broadest range of outcomes
   in {\it low-mass} massive stars because they are characterized by a very low envelope binding energy (Fig.~\ref{fig_eb_mr}).
   We indeed obtain light curves evolving from week to year time-scales (Fig.~\ref{fig_s11_lum_all})
   and ejecta expansion rates ranging from a few tens to a few thousand \kms (Fig.~\ref{fig_s11_phot}).
   An explosion/eruption producing a transient requires merely 10$^{49}$\,erg in such objects, a value
   that is so low that gravitational collapse of the stellar core may not be required.
   And indeed, in this mass range, stellar-evolutionary calculations reveal the existence of nuclear flashes
   in the last nuclear-burning stages \citep{weaver_woosley_79}, which could represent such an energy source.
   We therefore propose that low-mass massive stars are prime candidates for transient phenomena
   in the Universe, as well as prime canditates for interacting SNe such as 1994W \citep{dessart_etal_09}.

   In our simulations, sudden energy deposition above the core leads to shock-heating of the {\it entire} envelope. Whenever the
   energy deposited is greater than its binding energy, the {\it entire} envelope is ejected, with an asymptotic kinetic
   energy that is commensurate with the energy deposited.
   If a subsequent energy deposition occurs (e.g. a nuclear flash followed by gravitational collapse, as needed
   in a fraction at least of interacting SNe), the second ejection would have low mass and little or no hydrogen. Depending
   on the time between the two ejections, one could get an interacting SN for a short delay (i.e. a few years) or two transients
   at the same location for a long delay (the first one being dim if the explosion energy is small and the second one being potentially very dim
   due to the low ejected mass). These scenarios are rather speculative, but they are warranted since at present most, if not all, transients have
   an unknown origin and are poorly characterized.

    In our simulations, and within the context of this work,
   we obtain small mass ejections only when depositing the energy close to the progenitor surface,
   an eventuality that seems difficult to justify physically in such massive stars. Such low-mass ejections would seem
   to be better suited, for example, to the surface layers of an extended white dwarf. Observationally,
   low-mass ejections are likely to be associated with fast transients. For transients that are both fast and faint, a low-mass
   ejection in a highly- or moderately-bound object seems required. At the least, our simulations for (loosely-bound)
   RSG stars perturbed by a small deeply-rooted energy release produce large mass ejections (the overlying hydrogen
   envelope) and long-faint transients.

   Synthetic spectra computed for a sequence of models with varying energy deposition reveal a continuous
   evolution from Type II-P SN-like spectra at high energy ($L\sim$10$^8$\,\lsun), to low-luminosity Type II-P SN
   spectra at intermediate energy ($L\sim$10$^7$\,\lsun), to SN-impostor-like spectra at low energy  ($L\sim$10$^6$\,\lsun),
   with, in the same order, narrower line profiles and redder/cooler spectral-energy distributions (Fig.~\ref{fig_v1d_sed}).

   The results from this work should not be compromised by the approximations made in our radiation-hydrodynamics
   simulations.
   First, with one dimensionality we prevent convective transport and any structure formation through, e.g.,
   Rayleigh-Taylor instabilities. This may alter the properties of models characterized by low expansion speeds
   (longer evolution time-scale); we thus plan to study this eventuality with 2D and 3D simulations in the future.
   Second, we deposit the energy at a large and fixed rate, independent of any feedback effects. In the case of nuclear
   flashes, such feedback effects could shut-off the burning prematurely. We are aware of this artifact and will attempt in
   future work to develop a more physically-consistent approach by investigating the conditions that may lead to shock formation, rather
   than assuming a setup that systematically leads to it. However, provided a shock forms, we think
   our results apply.
   Third, progenitor models may differ on a mass-by-mass comparison with other groups but the general trend of increasing
   binding energy with main sequence mass should hold.
   Fourth, one-group transport should be accurate enough since it has been shown to capture the essentials of
   such radiation hydrodynamics simulations \citep{utrobin_chugai_09} - the key physics presented here takes place
   at large optical depth, under LTE conditions.

  Our finding that very modest energy perturbations can dramatically affect the structure of a massive star motivates
  detailed multi-dimensional hydrodynamical investigations of massive-star interiors, in particular of the last burning
  stages preceding core-collapse (see, e.g., \citealt{bazan_arnett_94,bazan_arnett_98,asida_arnett_00,
  arnett_etal_05,meakin_arnett_06}).
  As shown in these hydrodynamical simulations, and more generally,
  we surmise that the quasi-steady state approach of standard stellar-evolutionary codes (which keep an eye on the
  longer-term evolution of stars) may be missing important ingredients for our understanding of massive-star evolution.
  This has relevance for understanding mass loss and in particular massive-star eruptions, stellar variability/stability,
  and interacting SNe.
  Such pre-SN mass ejections would also modify, and perhaps sizably, the envelope mass and structure, thereby affecting
  the conditions surrounding core collapse and explosion.

    Ongoing and forthcoming surveys/missions like Pan-STARRS, Sky Mapper, the Palomar Transient Factory, GALEX,
    or the Large Synoptic Survey Telescope will better reveal the diversity of explosions, eruptions, and more generally
    transient phenomena in the Universe.
    We surmise that surveys up to now have missed a large number of low-energy long-lived transients, such as low-luminosity
    Type II-P SNe (objects even less energetic than SN1999br) and SN impostors. It is somewhat surprising that
    we have not yet detected Type II SNe with Plateau durations well in excess of 100 days.
   Moreover, for the shock-heating solutions presented here, a breakout signal systematically takes place.
   At SN-like energies, the signal may be too short to be resolved \citep{gezari_etal_08}, but for lower-energy transients,
   the reduced shock speed and strength would lengthen the breakout duration up to about a day, and
   move the peak of the spectral-energy distribution from $\sim$100\AA\ to the 1000-3000\AA\ range.
   Hence, the breakout signal should be more easily detectable in such transients, allowing
   to distinguish between an explosive event and, e.g., a super-Eddington wind.

\section*{Acknowledgments}

   LD acknowledges financial support from the European Community through an
   International Re-integration Grant, under grant number PIRG04-GA-2008-239184.


\end{document}